\RequirePackage{fix-cm}
\documentclass[12pt]{iopart}

\usepackage{amsgen}
\usepackage{amssymb}
\usepackage{setstack}
\usepackage{bbold}
\usepackage{mathrsfs}
\usepackage{graphicx}
\usepackage{epsfig}
\usepackage{epstopdf}
\usepackage{cite}
\usepackage{color}
\usepackage{rotating}

\def\be{\begin{equation}}
\def\ee{\end{equation}}
\def\bea{\begin{eqnarray}}
\def\eea{\end{eqnarray}}

\newcommand{\ba}{\begin{eqnarray}}
\newcommand{\ea}{\end{eqnarray}}
\newcommand{\dd}{{\rm{d}}}
\newcommand{\KK}{{\cal{K}}}
\newcommand{\HH}{{\cal{H}}}
\newcommand{\QQ}{{\cal{Q}}}
\newcommand{\DD}{{\cal{D}}}
\newcommand{\VV}{{\cal{V}}}
\newcommand{\RR}{{}^{(3)}{\cal{R}}}

\newcommand{\FF}{{\cal{F}}}
\newcommand{\GG}{{\cal{G}}}

\newcommand{\barOmmi}{\bar\Omega_{0} ^{(m)}}
\newcommand{\barOmki}{\bar\Omega_{0} ^{(k)}}

\newcommand{\Ommki}{\Omega_0 ^{(k)}}
\newcommand{\Ommmi}{\Omega_0 ^{(m)}}

\newcommand{\Omki}{\Omega_{q0} ^{(k)}}
\newcommand{\Ommi}{\Omega_{q0} ^{(m)}}

\newcommand{\ddrho}{\delta^{(m)}}
\newcommand{\ddKK}{\delta^{(k)}}

\newcommand{\bRR}{{}^{(3)}\bar{\cal{R}}}

\begin{document}

\title[Backreaction in Spherical and Plane Symmetric Dust-Filled Space-Times]{Cosmological Backreaction in Spherical and Plane Symmetric Dust-Filled Space-Times}

\author{Timothy Clifton$^1$}
\address{School of Physics \& Astronomy, Queen Mary University of London, UK.}
\vspace{-0.3cm}
\author{Roberto A. Sussman$^2$}
\address{Instituto de Ciencias Nucleares, Universidad Nacional Aut\'onoma de M\'exico (ICN-UNAM), A. P. 70--543, 04510 M\'exico D. F., M\'exico.}
\ead{$^1$t.clifton@qmul.ac.uk, $^2$sussman@nucleares.unam.mx}

\begin{abstract}
We examine the implementation of Buchert's and Green \& Wald's averaging formalisms in exact spherically symmetric and plane symmetric dust-filled cosmological models. We find that, given a cosmological space-time, Buchert's averaging scheme gives a faithful way of interpreting the large-scale expansion of space, and explicit terms that precisely quantify deviations from the behaviour expected from the Friedmann equations of homogeneous and isotropic cosmological models. The Green \& Wald formalism, on the other hand, does not appear to yield any information about the large-scale properties of a given inhomogeneous space-time. Instead, this formalism is designed to calculate the back-reaction effects of short-wavelength fluctuations around a given ``background'' geometry. We find that the inferred expansion of space in this approach is entirely dependent on the choice of this background, which is not uniquely specified for any given inhomogeneous space-time, and that the ``back-reaction'' from small-scale structures vanishes in every case we study. This would appear to limit the applicability of Green \& Wald's formalism to the study of large-scale expansion in the real Universe, which also has no pre-defined background. Further study is required to enhance the evaluation and comparison of these averaging formalisms, and determine whether the same difficulties exist, in less idealized space-time geometries.
\end{abstract}

\vspace{-0.5cm}
\section{Introduction}

Despite much work over the past decade, it remains an open question in cosmology as to whether small-scale inhomogeneities can have a sizable influence on the large-scale expansion of the Universe, or whether the expansion is always well described by the Friedmann solutions of Einstein's equations (with a suitably averaged energy-momentum content). Answering this question is of crucial importance for interpreting cosmological observations, and establishing the foundations of the cosmological models that are routinely used by observational cosmologists. However, despite this importance, there is as yet no universally accepted framework for determining either the large-scale expansion of the Universe, or the ``back-reaction'' effect that small-scale inhomogeneities have on that large-scale expansion.

Two leading contenders for how to average and calculate back-reaction in cosmology are those of Buchert \cite{buchert}, and Green \& Wald \cite{gw}. The two approaches prescribed in these two formalisms are quite different from each other, and it is of some interest to apply them to specific space-times in order to understand the degree to which they agree, or disagree. The Buchert formalism is based on a $1$+$3$-decomposition of the space-time, and an averaging of the scalar quantities involved in the Hamiltonian constraint and Raychaudhuri Equations over the $3$-spaces that result. The formalism developed by Green \& Wald, on the other hand, splits the metric of space-time into a ``background'' part and a ``perturbation'', and takes a limit in which the spatial scale and amplitude of the perturbations both reduce to zero at the same rate.

In this paper we will apply both the Buchert and Green \& Wald formalisms to a wide array of exact spherically symmetric and plane symmetric dust-filled space-times. These space-times will take the form of models that have locally homogeneous vacuum regions sandwiched between regions of locally homogeneous dust, as well as models in which the energy density of dust oscillates around a smooth background value. We will show that the two formalisms under investigation produce quite different results in a number of different models. This exemplifies the fact that these two formalisms are quantifying two quite different aspects of inhomogeneity. The precise aspects of back-reaction that they involve therefore need to be properly identified if the general back-reaction problem is to be fully understood, and we expect our treatment of this problem in the context of exact solutions to contribute to this understanding. This is a particularly interesting problem, given the controversy on the different interpretations of cosmological back-reaction that follow from the formalisms by Green \& Wald and Buchert (see References \cite{buchert2,buchert3,gw3,gw4,controversy1,controversy2}). 

The cosmological solutions we wish to study here are those that contain dust with $T^{ab}=\rho u^a u^b$, and which exhibit spherical or plane symmetry. These space-times have a line-element that can be written in the form \cite{exactsols}
\be \label{LTB1}
ds^2 = -e^{2 \nu (t)} \dd t ^2 + e^{2 \lambda (r,t)} \dd r^2 + R^2(r,t) \left[ \dd \theta^2 + \Sigma^2 (\theta , k) \dd \phi^2 \right]
\ee
where
\be
\Sigma (\theta, k) = \{ \sin \theta , \theta \} \qquad {\rm for} \qquad k = \{1,0\} \, .
\ee
This line-element contains all of the solutions we will study in this paper. It corresponds to spherical symmetry when $k=1$, and plane symmetry when $k=0$. We will start in Section \ref{hdm} by considering space-times that admit regions of vacuum sandwiched between regions of homogeneous dust, before proceeding to consider models in which the energy density of the dust is allowed to be inhomogeneous in Section \ref{idm}. 

There are two classes of solutions that take the form of Equation (\ref{LTB1}); those with $R'=0$, and those with $R' \neq 0$, where ${}'$ denotes $\partial/\partial r$. Let us now consider each of these two classes in turn:\\ \\

\hspace{-30pt}
{\it (i) Models with $R'=0$}: The vacuum space-times with $R'=0$ are either the $T$-region of Schwarzschild if $k=1$, or the degenerate Kasner solution if $k=0$. These can be written together with all non-vacuum dust solutions in this class if we first define the time coordinate such that $R(t)=t$, and then write the metric coefficients as
\begin{eqnarray}
e^{2 \nu(t)} ={\left(\frac{a}{t}-k\right)^{-1}} \qquad {\rm and} \qquad e^{\lambda(r,t)} = \left[ A(r) + B(r) V(t) \right] e^{-\nu(t)} \, ,
\end{eqnarray}
where $a$ is an arbitrary constant, $A(r)$ and $B(r)$ are arbitrary functions or $r$, and where $V(t)$ can be written as
\be
\label{pfield}
V(t) = \int^{e^{\nu}} \frac{2 x^2 dx}{1+k x^2} \,.
\ee
The energy density of dust in this class is then given by
\be
4 \pi \rho(r,t) = \frac{\dot{\lambda}(r,t) + \dot{\nu}(t)}{t\, e^{\nu(t)}} \, ,
\ee
where an over-dot denotes $\partial/\partial t$. It can be seen from this equation that the vacuum solutions will be recovered when $B(r)=0$, and that homogeneous solutions are recovered when $A(r)=0$. When $\rho \neq 0$ and $k=1$ these solutions are often thought of as generalizations of the homogeneous Kantowski-Sachs solution, while $\rho \neq 0$ and $k=0$ can be thought of as generalizations of Bianchi I space-times. This is the family of solutions that will be studied in Section \ref{hdm}.\\

\hspace{-30pt}
{\it (ii) Models with $R'\neq0$}: In the case the vacuum solutions are either given by the $R$-region of Schwarzschild if $k=1$, or by the vacuum Taub solution if $k=0$. They can be written together with all dust-filled solutions in this class if we choose a comoving frame with $u^a=\delta^a_t$, so that the metric coefficients in Equation (\ref{LTB1}) take the form
\begin{eqnarray}
\label{thing1}
e^{\nu(t)} = 1 \qquad {\rm and} \qquad e^{2 \lambda(r,t)} = \frac{R^{\prime 2}}{k-K(r)} \, ,
\end{eqnarray}
where $K=K(r)$ is an arbitary function, and where $R(r,t)$ must satisfy the following differential Equation:
\be
\label{Rdiff}
\dot{R}^2 = \frac{2 M}{R} - K(r) \, ,
\ee
where $M=M(r)$ is a second arbitrary function that can be used to write the energy density in dust in the form
\be
\label{thing2}
4 \pi G \rho = \frac{M'}{ R^2 R'} \, .
\ee
The solutions of Equation (\ref{Rdiff}) are well known, and on integration they introduce an additional free function $t_0(r)$, which is often referred to as the ``bang time'' (though the freedom to relabel the radial coordinate implies that only two free functions of $r$ are needed to specify a model). The vacuum solutions can be seen to be recovered in the case where $M$ is a constant. These solutions are often used as inhomogeneous generalizations and ``exact'' perturbations \cite{qaverage2,exactperts} of the dust-filled FLRW models. This is the family of solutions that will be studied in Section \ref{idm}.

\section{Two approaches to backreaction}

In this section we will introduce the Buchert formalism for scalar averaging and back-reaction, and then the Green \& Wald formalism for calculating the gravitational effects of short-wavelength, high-frequency perturbations. These two formalisms will then be applied to the exact solutions discussed above.

\subsection{The Buchert averaging prescription}

Buchert's equations are found by determining the expansion rate of a region of space, $\mathcal{D}$, and by averaging the energy density, scalar spatial curvature, and kinematic quantities over this domain. The results are \cite{buchert}
\begin{eqnarray}
3 \frac{\dot{a}^2_{\mathcal{D}}}{a_{\mathcal{D}}^2} = 8 \pi G_N \langle \rho \rangle_{\mathcal{D}} - \frac{1}{2} \left\langle^{(3)} R \right\rangle_{\mathcal{D}} - \frac{1}{2} \mathcal{Q}_{\mathcal{D}}
\label{buchert1}
\\[5pt]
3 \frac{\ddot{a}_{\mathcal{D}}}{a_{\mathcal{D}}} = - 4 \pi G_N \langle \rho \rangle_{\mathcal{D}} + \mathcal{Q}_{\mathcal{D}}
\label{buchert2}
\\[5pt]
\partial_t \langle \rho \rangle_{\mathcal{D}} + 3 \frac{\dot{a}_{\mathcal{D}}}{a_{\mathcal{D}}} \langle \rho \rangle_{\mathcal{D}} = 0,
\label{buchert3}
\end{eqnarray}
where 
\begin{eqnarray}
a_{\mathcal{D}}(t) &=& \left( \frac{\int_{\mathcal{D}} d^3 X \sqrt{\;^{(3)}g(t,X^i)}}{\int_{\mathcal{D}} d^3 X \sqrt{\;^{(3)}g(t_0,X^i)}} \right)^{\frac{1}{3}}
\\[5pt]
\left\langle \psi \right\rangle_{\mathcal{D}} &=& \frac{\int_{\mathcal{D}} d^3 X \psi (t,X^i) \sqrt{\;^{(3)}g(t,X^i)}}{\int_{\mathcal{D}} d^3 X \sqrt{\;^{(3)}g(t,X^i)}} \label{bav}
\\[5pt]
\mathcal{Q}_{\mathcal{D}} &=& \frac{2}{3} \left( \left\langle \Theta^2 \right\rangle_{\mathcal{D}} - \left\langle \Theta \right\rangle^2_{\mathcal{D}} \right) - 2 \left\langle \sigma^2 \right\rangle_{\mathcal{D}} \, ,
\label{Q}
\end{eqnarray}
and where the term $\mathcal{Q}_{\mathcal{D}}$ above is the kinematic backreaction. The scalar $\mathcal{Q}_{\mathcal{D}}$ gives us the back-reaction in the averaged Raychaudhuri equation (\ref{buchert2}), and is the quantity that we will calculate in the models that follow.

\subsection{The Green--Wald formalism}
\label{secgw}

In a series of articles, Green \& Wald (GW) developed a formalism to examine the back--reaction effect of small-scale (short-wavelength) inhomogeneities on the large-scale Universe. Their formalism is based on a series of four postulates \cite{gw}, which we will now summarize:
\\

\hspace{-30pt}
{\it Postulate 1}: Let $g_{ab}=g_{ab}(x^c,\lambda)$, with $\lambda>0$ be a one--parameter family of metrics that satisfies Einstein's field equations for all $\lambda>0$, such that
\begin{equation}
G_{ab}[g_{cd}(x^e,\lambda)]+8\pi g_{ab}(x^e,\lambda)\Lambda=8\pi T_{ab}(x^e,\lambda) \, ,\label{GW1}
\end{equation}
where $T_{ab}$ satisfies the weak energy condition. 
\\

\hspace{-30pt}
{\it Postulate 2}: There exist a bounded scalar function $C_1(x^a)$ such that for all $x^c$ we have
\begin{equation}
|\gamma_{ab}(x^c,\lambda)|\leq \lambda C_1,\qquad \gamma_{ab}\equiv g_{ab}(x^c,\lambda)-g_{ab}^{(0)}(x^c),\label{GW2}\end{equation}
where $g_{ab}^{(0)}(x^c)=\lim_{\lambda\to 0} g_{ab}(x^c,\lambda)$, and such that $g_{ab}^{(0)}$ need not satisfy Einstein's equations for the same $T_{ab}$ in Equation (\ref{GW1}).
\\

\hspace{-30pt}
{\it Postulate 3}: There exist a bounded scalar function $C_2(x^a)$ such that for all $x^c$  we have
\begin{equation} 
|\nabla_c \gamma_{ab}|\leq C_2,\label{GW3}
\end{equation}
where $\nabla_c$ is the covariant differential operator for the metric $g_{ab}^{(0)}$.
\\

\hspace{-30pt}
{\it Postulate 4}: For $h_{ab}$ and $\nabla_c h_{ab}$ there exists a smooth tensor field $\mu_{abcdef}$ such that for all smooth tensor fields $f^{abcdef}$ we have
\ba   \hbox{w-}\lim_{\lambda\to 0} \nabla_a \gamma_{cd}\nabla_b \gamma_{ef}
=\mu_{abcdef} , \label{GW4}\ea
where ``$\hbox{w-}\lim$'' denotes a weak limit.
\\

\noindent
If these postulates are valid, then Green \& Wald show that the ``averaged'' field equations can be written as
\begin{equation}
G_{ab}[g_{cd}^{(0)}]+8\pi g_{ab}^{(0)}\Lambda=8\pi T_{ab}^{(0)}+8\pi t_{ab}^{(0)},\label{GW5}
\end{equation}
where $T_{ab}^{(0)} = \hbox{w-}\lim_{\lambda\to 0} T_{ab}(\lambda)$, and where
\begin{eqnarray}
     t_{ab}^{(0)} &=& \frac{1}{8}\left[-\mu^c\,_c{}^{de}\,_{de}-\mu^c\,_c{}^{d}\,_{d}{}^{e}\,_{e}+2\mu^{cd}\,_c{}^e\,_{de}\right]\,g_{ab}^{(0)}
     +\frac{1}{2}\mu^{cd}\,_{acbd}-\frac{1}{2}\mu^c\,_{ca}{}^d\,_{bd}     \nonumber\\
 &&\qquad +\frac{1}{4}\mu_{ab}{}^{cd}\,_{cd}-\frac{1}{2}\mu^c\,_{(ab)c}{}^d\,_d+\frac{3}{4}\mu^c\,_{cab}{}^d\,_d-\frac{1}{2}\mu^{cd}\,_{abcd},\label{GW7}
\end{eqnarray}
with the tensor $t_{ab}^{(0)}$ satisfying $[t^{(0)}]^a\,_a=0$ and $8\pi t_{ab}^{(0)}t^at^b\geq 0$ for all timelike fields $t^a$. The tensor $t_{ab}^{(0)}$, in this approach, quantifies the back-reaction effect that the small-scale inhomogeneities have on the large-scale expansion of space. This approach is a generalization of the Isaacson averaging scheme \cite{isaac1,isaac2}, which was originally intended to quantify the gravitational field of short-wavelength gravitational waves, and which was formulated in terms of weak limits by Burnett \cite{burnett}.

\section{Homogeneous dust and vacuum models}
\label{hdm}

\begin{figure}
\begin{center}
\includegraphics[scale=1]{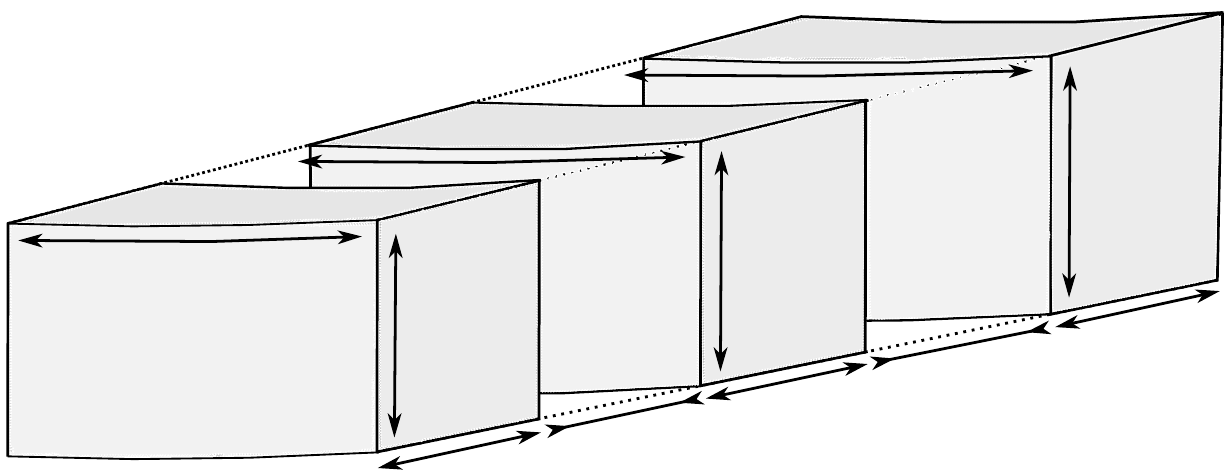}
\caption{An illustration of the situation studied in Section \ref{hdm}. Shaded regions represent the homogeneous dust-filled regions, which sandwich regions of homogeneous vacuum. The configuration is intended to be extended in each spatial dimension {\it ad infinitum}.}
\label{slabs}
\end{center}
\end{figure}

The first class of models we wish to consider are those in which regions of homogeneous vacuum are sandwiched between regions of homogeneous dust, as depicted in Figure \ref{slabs}. This can be achieved by considering the solutions with $R'=0$, and will be done first of all for plane-symmetric space-times (where $k=0$), and then for spherically symmetric space-times (where $k=1$).

\subsection{Plane symmetry}

If $R'=0$ and $k=0$ then the integration in Equation (\ref{pfield}) can be performed straighforwardly, and the time coordinate can be re-defined so that the line-element in Equation (\ref{LTB1}) becomes
\be \label{sliceds}
ds^2 = -\dd t ^2 + \left[ c_1 \, \left(\frac{t}{t_0}\right)^{\frac{2}{3}} + c_2 \left(\frac{t_0}{t}\right)^{\frac{1}{3}} \right]^2 \dd r^2 + \left(\frac{t}{t_0}\right)^\frac{4}{3} \left( \dd \theta^2 + \theta^2 \dd \phi^2 \right) \, ,
\ee
where $c_1=c_1(r)$ and $c_2=c_2(r)$ are arbitrary functions of $r$, and $t_0$ is a constant. In this case, the energy density can be written
\be
\rho = \frac{c_1}{6 \pi t (c_1 \, t+c_2 \, t_0)} \, .
\ee
This is the general plane-symmetric dust filled solution with $R^{\prime}=0$ and $\dot{R} \neq 0$, and can be seen to reduce to the Einstein-de Sitter (EdS) when $c_2=0$, and the degenerate Kasner solution when $c_1=0$. In what follows, we choose units such that $t_0=1$.

We now wish to consider situations in which we have slices of Einstein-de Sitter geometry sandwiched between regions of Kasner vacuum, repeated over and over again forever. This is a special case of the geometry in Equation (\ref{sliceds}), which can be shown to explicitly satisfy the required junction conditions between neighbouring regions of dust and vacuum, and hence constitute a viable family of cosmological solution to Einstein's equations \cite{slice1,slice2,slice3}. Although they are too symmetric to describe any realistic astrophysical structures, they do provide an interesting framework to explore ideas about inhomogeneity and anisotropy in the context of exact solutions. More specifically, they have been shown to have interesting, non-trivial properties when averaged \cite{slice4}.\\

\newpage

\noindent
{\it Buchert Averages:}
\\

\noindent
In the regions of Kasner space (where $c_1=0$) the expansion and shear scalars are given by the following expressions:
\be \label{kvk}
\Theta_{\rm K} = \frac{1}{t} \qquad {\rm and} \qquad \sigma^2_{\rm K} = \frac{1}{3 t^2} \, ,
\ee
where $\Theta \equiv D_a u^a$ is the expansion rate of the time-like geodesics with tangent vector $u^a$ that stay at fixed values of the spatial coordinates $\{r, \theta, \phi\}$, and where $\sigma^2 \equiv \frac{1}{2} \sigma_{ab} \sigma^{ab}$ is the magnitude of the shear tensor $\sigma_{ab} \equiv D_{(a} u_{b)} - \frac{1}{3} h_{ab} \, D_c u^{c}$. The projection tensor in these expressions is defined as $h_{ab} \equiv g_{ab} + u_a u_b$, and the derivative operator is projected so that (for example) $D_{a} u_{b} = h_a^{\phantom{a} c} h_b^{\phantom{b} d} \nabla_c u_d$.

Correspondingly, the expansion and shear scalars in the region of Einstein-de Sitter space (where $c_2=0$) are given by
\be \label{kveds}
\Theta_{\rm EdS} = 3 H= \frac{2}{t} \qquad {\rm and} \qquad \sigma^2_{\rm EdS} = 0 \, ,
\ee
where $H=\dot{a}/{a}$ is the Hubble expansion rate of these regions, and where we have again chosen the time-like geodesics curves at constant values of $\{r, \theta, \phi\}$ to define these kinematic variables. The result $\sigma^2_{\rm EdS} = 0$ follows immediately from the fact that these regions of space are locally isotropic. The same time coordinate can be used in both Equations (\ref{kvk}) and (\ref{kveds}), as the union of the sets of geodesic curves at constant values of $\{r, \theta, \phi\}$ in both types of regions together form a congruence that threads the entire cosmology.

If we now choose an averaging domain that is larger than the homogeneity scale of this space-time, such that it encompasses exactly one region of EdS and one region of Kasner space, then we can calculate the consequences of the inhomogeneity and isotropy of these models on the large-scale course-grained expansion. Taking the spatial extent of the Kasner region to be given by the coordinate separation $\Delta r_{\rm K}$, and the extent of the EdS regions to be $\Delta r_{\rm EdS}$, we find that the Buchert averaged expansion and shear of our domains to be given by Equation (\ref{bav}) as
\ba
\langle \Theta \rangle &=& \frac{2 t + \eta}{t (t + \eta)} \\[5pt]
\langle \Theta^2 \rangle &=& \frac{4 t + \eta}{t^2 (t + \eta)} \\[5pt]
\langle \sigma^2 \rangle &=& \frac{\eta}{3 t^2 (t + \eta)} \, ,
\ea
where $\eta \equiv \Delta r_{\rm K}/\Delta r_{\rm EdS}$. Subsituting these values into the expression for the back-reaction scalar in Equation (\ref{Q}) gives the simple result
\be
Q = - \frac{2 \eta^2}{3 t^2 (t+\eta)^2} \, ,
\ee
which can be seen have the following early and late-time limits:
\be
\lim_{t \rightarrow 0} Q = - \lim_{t \rightarrow 0} \frac{2}{3 t^2} = - \infty \qquad {\rm and} \qquad \lim_{t \rightarrow \infty} Q = - \lim_{t \rightarrow \infty} \frac{2 \eta^2}{3 t^4} = 0_- \, .
\ee
This result makes sense physically as the early-time behaviour is dominated by the anisotropic vacuum Kasner regions, while the late-time behaviour is dominated by the isotropic dust-filled EdS regions. The anisotropic early stage to these space-times should therefore be expected to give a large deviation from any naive expectations obtained from using the Friedmann equation, which is verified in this case by the divergence of the back-reaction scalar in the limit $t\rightarrow 0$. 

The Buchert averaging and back-reaction scheme appears to give simple results with direct physical interpretation: The averaged expansion and shear scalars express the behaviour of the integration domain we have chosen, and the back-reaction scalar $Q$ gives the extra term that should be included in an effective Friedmann equation for this domain if we are to correctly calculate the evolution of its spatial volume.\\

\noindent
{\it Green \& Wald Averages:}
\\

\noindent
We now wish to analyse the behaviour of these exact same space-times within the framework developed by Green \& Wald. This requires us to split our metric into two parts, so that
\be
g_{ab} = g^{(0)}_{ab} + \gamma_{ab} \, ,
\ee
where there is no approximation or perturbation scheme involved, but where $g^{(0)}_{ab}$ is considered as a ``background'' about which $\gamma_{ab}$ fluctuates. This presents a problem in the present case, as there does not appear to be a unique presciption for performing this split. We will therefore consider several different possibilities.\\

\noindent
{\it Case I: An Einstein-de Sitter background.} The first possibility we wish to study is one in which the EdS geometry is taken as the background, and the vacuum regions are considered as fluctuations around this background. The background can then be specified by choosing $c_1=1$ and $c_2=0$ in Equation (\ref{sliceds}), which gives the expected geometry:
\be
\label{eds}
g^{(0)}_{ab} dx^a dx^b = -dt^2 + t^{\frac{4}{3}} d {\bf x}^2 \, ,
\ee
where $d {\bf x}^2$ is the line-element of flat three-dimensional Euclidean space. The regions of Kasner space, which are treated as fluctuations around this background, are then considered to be exact metric perturbations. 

We choose the Einstein-de Sitter regions and Kasner regions to both have width $r_*$, such that $\Delta r_K = \Delta r_{EdS} =r_*$. We will do this, without loss of generality, by taking the region $0 < r < r_*$ to be Kasner space, and the region $r_*< r < 2r_*$ to be EdS. These two regions will then be followed by alternating regions of Kasner and EdS that continue forever, to create a global cosmological model. In the Kasner region ($0 < r < r_*$) we can then take the values of $c_1$ and $c_2$ to be $c_1=0$ and $c_2=\frac{1}{2} \left[ \cos \left\{ 2 \pi \left( \frac{r}{r_*}-\frac{1}{2} \right) \right\} +1 \right]$, while in the EdS region ($r_*< r < 2r_*$) we can take $c_1=\frac{1}{2} \left[ \cos \left\{ 2 \pi \left( \frac{r}{r_*}-\frac{1}{2} \right) \right\} +1 \right]$ and $c_2=0$.

\begin{figure}
\begin{center}
\includegraphics[scale=1]{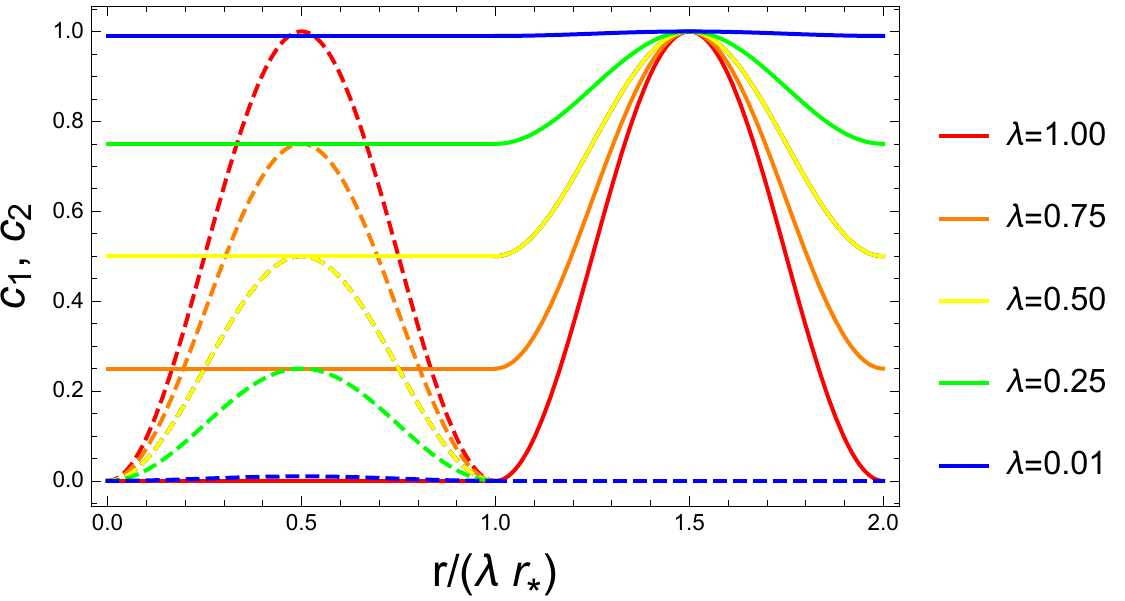}
\caption{The values of $c_1$ (solid lines) and $c_2$ (dashed lines) for models with an EdS background.}
\label{c1c2a}
\end{center}
\end{figure}

We must now choose a way to parameterize this geometry with $\lambda$ such that when $\lambda \rightarrow 0$ we recover the ``background'' geometry given in Equation (\ref{eds}). There is no unique prescription for doing this provided by Green \& Wald, so we make the following choices: In the regions of Kasner space we choose
\begin{eqnarray}
c_{1}^{K} &=& 1 - \lambda \\
c_{2}^{K} &=& \frac{\lambda}{2} \left[ \cos \left\{ 2 \pi \left( \frac{r}{\lambda r_*}-\frac{1}{2} \right) \right\} +1 \right] \, ,
\end{eqnarray}
while in the regions of EdS we choose
\begin{eqnarray}
c_{1}^{EdS} &=& 1+ \frac{\lambda}{2} \left[ \cos \left\{ 2 \pi \left( \frac{r}{\lambda r_*}-\frac{1}{2} \right) \right\} -1 \right] \\
c_{2}^{EdS} &=& 0 \, .
\end{eqnarray}
We have in mind here that the width of the regions that are initially Kasner space are also parameterized to be $\Delta r_K = \lambda r_*$, while the regions that are initially EdS have width $\Delta r_{EdS} = \lambda r_*$. These choices are shown graphically in Figure \ref{c1c2a}, and can be seen to interpolate between the inhomogeneous macroscopic space-time when $\lambda=1$, and the pure EdS background in the limit $\lambda \rightarrow 0$. This paramaterization has been chosen to explicitly satisfy Green \& Wald's postulates, while keeping $\nabla_c \gamma_{ab} \neq 0$ in the limit $\lambda \rightarrow 0$, and hence allowing for the possibility of a non-zero $\mu_{abcdef}$.

The presciption set out above is an exact perturbation of EdS, with $\gamma_{ab}=g_{ab}-g^{(0)}_{ab}$ given by the following expression in the Kasner region ($0<r<\lambda r_*$):
\be \label{gameds1}
\hspace{-30pt}
\gamma^K_{ab} dx^a dx^b = \left[ 
\left( 1-\lambda  + \frac{\lambda}{2}\left[  \cos \left\{ 2 \pi \left( \frac{r}{\lambda r_*}-\frac{1}{2} \right) \right\} +1 \right] \frac{1}{t} \right)^2- 1
\right] t^{\frac{4}{3}} dr^2 \, ,
\ee
and the following expression in the EdS region ($\lambda r_*<r< 2 \lambda r_*$):
\be \label{gameds2}
\hspace{-30pt}
\gamma^{EdS}_{ab} dx^a dx^b = \left[ 
\left( 1 + \frac{\lambda}{2}\left[  \cos \left\{ 2 \pi \left( \frac{r}{\lambda r_*}-\frac{1}{2} \right) \right\} -1 \right] \right)^2 - 1
\right] t^{\frac{4}{3}} dr^2 \, .
\ee
The amplitude of these perturbations can be seen to reduce to zero in the limit $\lambda \rightarrow 0$, while the first derivative does not. This is the so-called ``high-frequency limit'', where the amplitude and spatial scale of the perturbations both shrink to zero simultaneously. The fluctuation above can be seen to satisfy Einstein's equations, as it gives a geometry that is of exactly the form of Equation (\ref{sliceds}) for all $0 < \lambda <1$. One can trivially arrange for this to be true for all positive values of some new parameter $\lambda^{\prime}$ if we define, for example, $\lambda^{\prime} = \tanh^{-1} \lambda$. Our fluctuation therefore obeys G\&W's Postulate 1. The form of Equations (\ref{gameds1})-(\ref{gameds2}) can also be seen to obey Postulate 2, as every term in each expression is proportional to either $\lambda$ or $\lambda^2$. Finally, explicit calculation shows that every component of $\nabla_c \gamma_{ab}$ is proportional to $\lambda$ or $\lambda^2$, except $\nabla_{r} \gamma_{rr}$ which is given by
\be \label{dgrrr1}
\nabla_r \gamma^K_{rr} = - \frac{2 \pi t^{\frac{1}{3}}}{r_*} \sin \left\{ 2 \pi  \left(\frac{r}{\lambda r_*}-\frac{1}{2} \right) \right\} + \mathcal{O}( \lambda) \, ,
\ee
and
\be \label{dgrrr2}
\nabla_r \gamma^{EdS}_{rr} = - \frac{2 \pi t^{\frac{4}{3}}}{r_*} \sin \left\{ 2 \pi  \left(\frac{r}{\lambda r_*}-\frac{1}{2} \right) \right\} + \mathcal{O}( \lambda) \, .
\ee
This shows that Postulate 3 is also satisfied. Substituting this result into the definition of $\mu_{abcdef}$ then gives
\begin{eqnarray}
\mu_{abcdef} &=& \frac{2 \pi^2 t^{\frac{2}{3}}}{r_*^2} \delta^r_a \delta^r_b\delta^r_c\delta^r_d\delta^r_e\delta^r_f  \qquad {\rm in \; the\; Kasner\; regions} \\
\mu_{abcdef} &=& \frac{2 \pi^2 t^{\frac{8}{3}}}{r_*^2} \delta^r_a \delta^r_b\delta^r_c\delta^r_d\delta^r_e\delta^r_f  \qquad {\rm in \; the\; EdS\; regions} \, .
\end{eqnarray}
In both cases it can be seen, from the definition of $t^{(0)}_{ab}$ in Equation (\ref{GW7}), that
\be
t^{(0)}_{ab}=0 \qquad {\rm for \; all} \; r \, .
\ee
The Kasner-EdS space-time can therefore be modelled, within the approach of G\&W, as an EdS background with zero back-reaction from the Kasner underdensities. Let us now consider other choices we could have made for the background part of the geometry.\\

\noindent
{\it Case II: A Kasner background.} We could equally well have chosen the background geometry $g^{(0)}_{ab}$ to be Kasner, such that $c_1=0$ and $c_2=1$ in Equation (\ref{sliceds}), in which case the background line-element reads
\be
\label{kas}
g^{(0)}_{ab} dx^a dx^b = -dt^2 + \frac{dr^2}{t^{\frac{2}{3}}} + t^{\frac{4}{3}} ( d\theta^2 + \theta^2 d \phi^2 )\, .
\ee
The regions of EdS space can now be treated as fluctuations around this Kasner background, with $c_1$ and $c_2$ given by
\begin{eqnarray}
c_{1}^{K} &=& 0  \\
c_{2}^{K} &=& 1 + \frac{\lambda}{2} \left[ \cos \left\{ 2 \pi \left( \frac{r}{\lambda r_*}-\frac{1}{2} \right) \right\} -1 \right] \, ,
\end{eqnarray}
while in the regions of EdS we choose
\begin{eqnarray}
c_{1}^{EdS} &=& \frac{\lambda}{2} \left[ \cos \left\{ 2 \pi \left( \frac{r}{\lambda r_*}-\frac{1}{2} \right) \right\} +1 \right]\\
c_{2}^{EdS} &=& 1-\lambda \, .
\end{eqnarray}
We have again chosen to take the Kasner region to occupy $0< r< \lambda r_*$, and the EdS region to occupy $\lambda r_* < r < 2 \lambda r_*$, with alternate such regions following on forever.

\begin{figure}
\begin{center}
\includegraphics[scale=1]{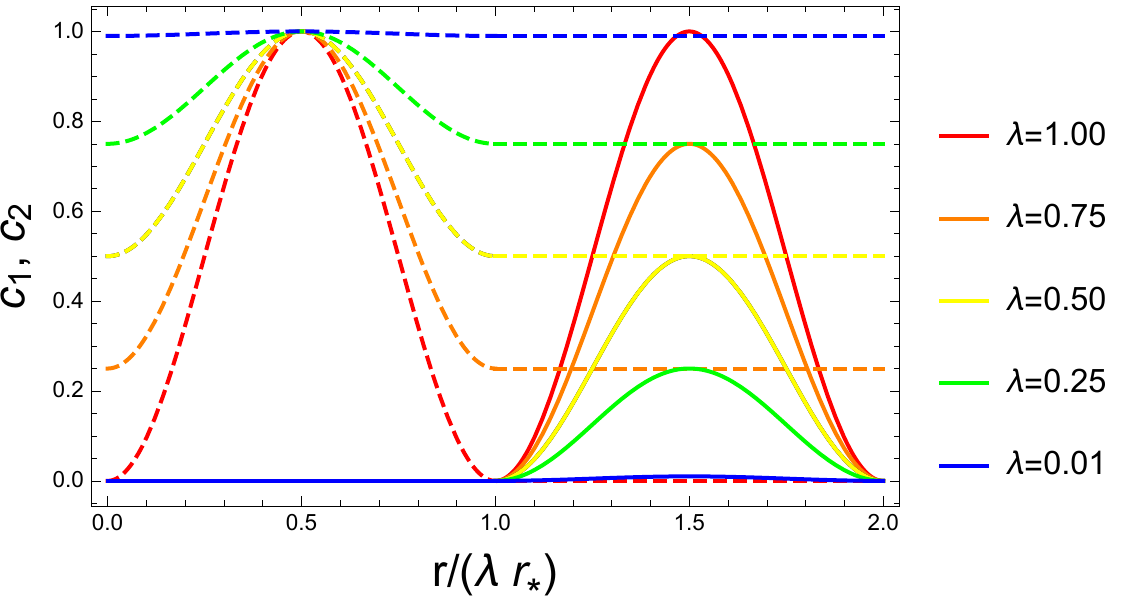}
\caption{The values of $c_1$ (solid lines) and $c_2$ (dashed lines) for models with a Kasner background.}
\label{c1c2b}
\end{center}
\end{figure}

These choices of $c_1$ and $c_2$ are illustrated in Figure \ref{c1c2b}, and lead to the following exact perturbations to the ``background'' geometry specified in Equation (\ref{kas}):
\be \label{gamkas1}
\hspace{-30pt}
\gamma_{ab}^K dx^a dx^b = \left[
\left( 1 + \frac{\lambda}{2}\left[  \cos \left\{ 2 \pi \left( \frac{r}{\lambda r_*}-\frac{1}{2} \right) \right\} -1 \right] \right)^2 - 1
\right] \frac{1}{t^{\frac{2}{3}}}\, dr^2
\ee
and
\be \label{gamkas2}
\hspace{-30pt}
\gamma_{ab}^{EdS} dx^a dx^b = \left[
\left( 1-\lambda + \frac{\lambda}{2}\left[  \cos \left\{ 2 \pi \left( \frac{r}{\lambda r_*}-\frac{1}{2} \right) \right\} +1 \right] t \right)^2 - 1
\right] \frac{1}{t^{\frac{2}{3}}}\, dr^2
\ee
where $\lambda \in ( 0,1 )$ is again the parameter used to take the high-frequency limit.

We now have that the perfect Kasner and EdS geometries are approached as $\lambda \rightarrow 1$, and that the fluctuation around the Kasner background is entirely removed in the limit $\lambda \rightarrow 0$. The fluctuation above can be seen to satisfy G\&W's Postulates 1, 2 and 3, with every component of $\nabla_c \gamma_{ab}$ being proportional to $\lambda$ or $\lambda^2$ except $\nabla_{r} \gamma_{rr}$, which is now given by 
\be \label{dgrrr1b}
\nabla_r \gamma^K_{rr} = - \frac{2 \pi}{r_* t^{\frac{2}{3}}} \sin \left\{ 2 \pi  \left(\frac{r}{\lambda r_*}-\frac{1}{2} \right) \right\} + \mathcal{O}( \lambda) \, ,
\ee
and
\be \label{dgrrr2b}
\nabla_r \gamma^{EdS}_{rr} = - \frac{2 \pi t^{\frac{1}{3}}}{r_*} \sin \left\{ 2 \pi  \left(\frac{r}{\lambda r_*}-\frac{1}{2} \right) \right\} + \mathcal{O}( \lambda) \, .
\ee
The values of $\mu_{abcdef}$ and $t^{(0)}_{ab}$ are then found to be
\begin{eqnarray}
\mu_{abcdef} &=& \frac{2 \pi^2}{r_*^2 t^{\frac{4}{3}}} \delta^r_a \delta^r_b\delta^r_c\delta^r_d\delta^r_e\delta^r_f  \qquad {\rm in \; the\; Kasner\; regions} \\
\mu_{abcdef} &=& \frac{2 \pi^2 t^{\frac{2}{3}}}{r_*^2} \delta^r_a \delta^r_b\delta^r_c\delta^r_d\delta^r_e\delta^r_f  \qquad {\rm in \; the\; EdS\; regions} \, ,
\end{eqnarray}
and
\be
t^{(0)}_{ab}=0 \qquad {\rm for \; all} \; r \, ,
\ee
which is exactly the same as when EdS was taken as the background, and again gives zero back-reaction. This is somewhat puzzling, as it appears that the Kasner-EdS geometry can equally well be described with either Kasner or EdS as the background, and that in both cases there is zero effect from inhomogeneities on the expansion of that background. This appears to be the case even though Kasner and EdS space-times expand quite differently from each other.\\

\begin{figure}
\begin{center}
\includegraphics[scale=1]{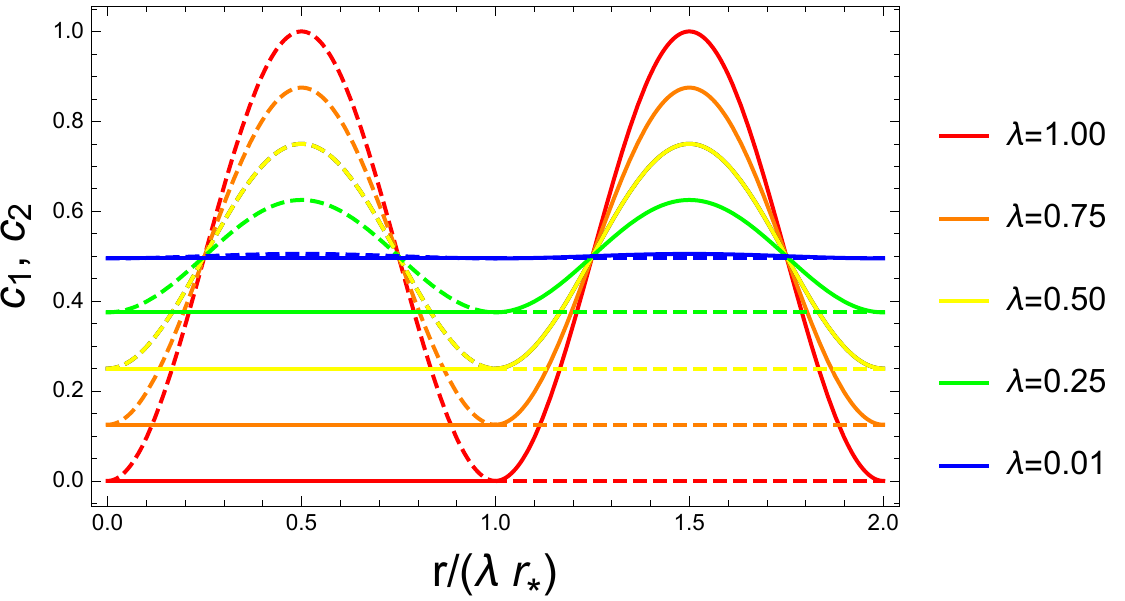}
\caption{The values of $c_1$ (solid lines) and $c_2$ (dashed lines) for models with a Bianchi type I dust-filled background.}
\label{c1c2c}
\end{center}
\end{figure}

\noindent
{\it Case III: A dust-filled Bianchi I background.}  Finally, we could take the background of the Kasner-EdS model to be a dust-filled space-time of Bianchi type I. This is closer to what would we expect most people would naively consider to be an averaged version of this space-time. In this case we can choose $c_1=c_2=1/2$ in Equation (\ref{sliceds}), to get the following background:
\be
g^{(0)}_{ab} dx^a dx^b = -dt^2 + \left( \frac{t^{\frac{2}{3}}}{2} +\frac{1}{2 t^{\frac{1}{3}}} \right)^2 {dr^2} + t^{\frac{4}{3}} ( d\theta^2 + \theta^2 d \phi^2 )\, .
\ee
We now need to treat both the Kasner regions and the EdS regions as fluctuations around this Bianchi type I space-time. Starting with $c_1$ and $c_2$, we can write
\begin{eqnarray}
c_{1}^{K} &=& \frac{1-\lambda}{2} \\
c_{2}^{K} &=& \frac{1}{2} + \frac{\lambda}{2}   \cos \left\{ 2 \pi \left( \frac{r}{\lambda r_*}-\frac{1}{2} \right) \right\}
\end{eqnarray}
while in the regions of EdS we choose
\begin{eqnarray}
c_{1}^{EdS} &=&  \frac{1}{2} + \frac{\lambda}{2}   \cos \left\{ 2 \pi \left( \frac{r}{\lambda r_*}-\frac{1}{2} \right) \right\} \\
c_{2}^{EdS} &=& \frac{1-\lambda}{2}
\end{eqnarray}
where once more the Kasner region has been chosen to occupy $0< r< \lambda r_*$, and the EdS region $\lambda r_* < r < 2 \lambda r_*$. These choices are displayed graphically in Figure \ref{c1c2c}.

Calculating the derivatives of $\gamma_{ab}=g^{(0)}_{ab}-g_{ab}$ we find that the only component containing any terms that are not directly proportional to either $\lambda$ or $\lambda^2$ is
\be \label{dgrrr1c}
\nabla_r \gamma^K_{rr} = - \frac{\pi (1+t)}{r_* t^{\frac{2}{3}}} \sin \left\{ 2 \pi  \left(\frac{r}{\lambda r_*}-\frac{1}{2} \right) \right\} + \mathcal{O}( \lambda) \, ,
\ee
and
\be \label{dgrrr2c}
\nabla_r \gamma^{EdS}_{rr} = - \frac{-\pi (1+t) t^{\frac{1}{3}}}{r_*} \sin \left\{ 2 \pi  \left(\frac{r}{\lambda r_*}-\frac{1}{2} \right) \right\} + \mathcal{O}( \lambda) \, .
\ee
Correspondingly we then find
\begin{eqnarray}
\mu_{abcdef} &=& \frac{\pi^2 (1+t)^2}{2 r_*^2 t^{\frac{4}{3}}} \delta^r_a \delta^r_b\delta^r_c\delta^r_d\delta^r_e\delta^r_f  \qquad {\rm in \; the\; Kasner\; regions} \\
\mu_{abcdef} &=& \frac{\pi^2 t^{\frac{2}{3}}(1+t)^2}{2 r_*^2} \delta^r_a \delta^r_b\delta^r_c\delta^r_d\delta^r_e\delta^r_f  \qquad {\rm in \; the\; EdS\; regions} \, ,
\end{eqnarray}
and
\be
t^{(0)}_{ab}=0 \qquad {\rm for \; all} \; r \, .
\ee
Again, there is no back-reaction. And again it appears that the background dust-filled Bianchi type I geometry is another acceptable background.

\subsection{Spherical symmetry}

If $R'=0$ and $k=1$ then the integration in Equation (\ref{pfield}) can still be performed, and is given parametrically by
\begin{eqnarray}
V &=& K+ N (\Psi +\cot \Psi)\\
t &=& a \cos^2 \Psi \,
\end{eqnarray}
where $K$, $N$ and $a$ are constants. The metric in Equation (\ref{LTB1}) can then be written
\be 
\hspace{-30pt}
ds^2 = -\frac{\dd t ^2}{\left(\frac{a}{t}-1\right)} +  \left( A(r) + B(r) V(t) \right) \left(\frac{a}{t}-1\right) \dd r^2 + t^2 \left( \dd \theta^2 + \sin^2 \theta \, \dd \phi^2 \right) \, ,
\ee
where $A(r)$ and $B(r)$ are arbitrary functions of $r$. This is the general spherically-symmetric dust-filled solution with $R^{\prime}=0$ and $\dot{R} \neq 0$. It reduces to the homogeneous Kantowski-Sachs solution when $A=0$, and is isometric to the T-region of Schwarzschild when $B=0$.

As in the previous section we wish to consider regions of vacuum sandwiched between regions of homogeneous dust. In the present case the vacuum will correspond to the Schwarzschild T-region, and the dust will correspond to the Kantowski-Sachs solution. As we move through space-time in the $r$ direction, we will alternate between these two geometries, and one can imagine a universe constructed from infinitely many of these regions joined together at suitable boundaries (again, as illustated in Figure \ref{slabs}). In the rest of this section we will investigate what the Buchert averaging scheme and the Green \& Wald averaging scheme can tell us about back-reaction and the large-scale properties of such a set up.\\

\noindent {\it Buchert Averages:}\\

\noindent Let us first consider the Buchert average of the space-times constructed from sandwiching together vacuum Schwarzschild T-regions of coordinate width $\Delta r_{ST}$ and homogeneous dust-filled Kantowski-Sachs regions of width $\Delta r_{KS}$. The expansion and shear scalars for a set of geodesic observers, with constant $r$, $\theta$ and $\phi$ coordinates, can be calculated in each of these two regions. 

In the vacuum Schwarzschild T-regions, where $B=0$, we find these to be given by
\be
\Theta_{ST} = \frac{3 a -4 t}{2 \sqrt{a-t}\, t^{\frac{3}{2}}} \qquad {\rm and} \qquad \sigma^2_{ST} = \frac{(3a-2t)^2}{12 (a-t) t^3} \, .
\ee
On the other hand, in the Kantowski-Sachs dust-filled regions, where we can take $A=0$, we find
\be
\Theta_{KS} = \frac{(3 a-4 t) V +2 (a-t) V'}{2 \sqrt{a-t}\, t^{\frac{3}{2}} \, V}
\ee
and
\be
\sigma^2_{KS} = \frac{((3a-2t) V-2t(a-t) V')^2}{12 (a-t) t^3 V^2} \, .
\ee
From these two sets of scalars it is now straightforward to calculate their spatial average over a domain that contains one vacuum region and one dust-filled region, and that we therefore expect to give the global average expansion and shear of the entire space-time (as the space-time can be constructed from reproducing such regions over and over again, forever).

The average of the expansion scalar over such a domain is given by
\be
\langle \Theta \rangle = \frac{(3a-4t) (\eta +V) +2 (a-t) t V'}{2 \sqrt{a-t} \, t^{\frac{3}{2}} \, (\eta+V)} \, ,
\ee
while the average of the expansion scalar squared is given by
\be
\langle \Theta^2 \rangle = \frac{(3a-4t)^2}{4 (a-t) t^3} + \frac{V' ((3a-4t) V +(a-t) t V')}{V (\eta+V) t^2} \, .
\ee
Finally, the average of the shear scalar can be found to be
\be
\langle \sigma^2 \rangle =\frac{(3a-2t)^2}{12 (a-t) t^3} - \frac{V' ((3a-2t) V -(a-t) t V')}{3 V (\eta+V) t^2} \, ,
\ee
where $\eta = \Delta r_{ST}/\Delta r_{KS}$. These quantities can be combined, as in Equation (\ref{Q}), to find the back-reaction scalar to be
\be \label{buchertq1}
Q= - \frac{((3a-2t) (\eta+V)-2 (a-t) t V')^2}{6 (a-t) t^3 (\eta+V)^2} \, .
\ee
This scalar is clearly non-zero, and gives the contribution to the effective Friedmann equation that is required to reproduce the average expansion of the spatial domain under consideration. When $\eta \rightarrow 0$ this gives the back-reaction scalar in a purely Kantowski-Sachs space-time, and when $\eta \rightarrow \infty$ it gives the corresponding quantity in the T-region of Schwarzschild. Both of these limits result in non-zero $Q$, as each of the two different regions in this case are anisotropic, and therefore have a different global expansion rate from that which would be prescribed by the Friedmann equation.\\

\noindent {\it Green \& Wald Averages:}\\

\noindent In trying to apply the Green \& Wald formalism to this class of space-times we are again forced to separate the metric into a ``background'' part and a ``perturbation''. As before, there is no prescription provided for how to do this, and so we are forced to make a choice as to what we consider the background to be. We will consider three such backgrounds: One given by the homogeneous dust-filled regions, a second given by the vacuum Schwarzschild T-regions, and a third given by a homogeneous intermediate geometry.\\

\noindent {\it Case IV: A Kantowski-Sachs background.} In this case we consider the dust regions to be the ``background'' geometry, and the vacuum regions to be exact perturbations about this background. This can be achieved by making the following choice for $A(r)$ and $B(r)$ in the vacuum region:
\be
A^{ST} = \frac{\lambda}{2} \left[ \cos \left\{ 2 \pi \left( \frac{r}{\lambda r_*} - \frac{1}{2} \right) \right\} +1 \right]
\ee
\be
B^{ST}=1-\lambda \, ,
\ee
while choosing the values of these functions in the dust-filled regions to be
\be
A^{KS}=0
\ee
\be
B^{KS}=1+\frac{\lambda}{2} \left[ \cos \left\{ 2 \pi \left( \frac{r}{\lambda r_*} - \frac{1}{2} \right) \right\} -1 \right] \, .
\ee
These functions are the same as the ones depicted in Figure \ref{c1c2a}, with $A \leftrightarrow c_2$ and $B \leftrightarrow c_1$, such that the Schwarzschild T-region is given by $0 <r < \lambda r_*$ and the Kantowsi-Sachs region is in the interval $ \lambda r_*< r < 2\lambda r_*$ (repeated over and over again, forever).

In each of the two regions the perturbation to the Kantowski-Sachs background (with $\lambda \rightarrow 0$) can then be written as
\be \label{KSSTpert}
\hspace{-30pt}
\gamma_{ab} dx^a dx^b = \left[
\left(\sqrt{\frac{a-t}{t}} (A+ B\,  V ) \right)^2 - \left( \sqrt{\frac{a-t}{t}} V \right)^2
\right] \, dr^2
\ee
which gives the only components of $\nabla_a \gamma_{bc}$ that are not proportional to $\lambda$ or $\lambda^2$ as being
\be
\nabla_r \gamma^{ST}_{rr} = \frac{2 \pi (t-a) V}{r_* \, t} \sin \left\{ 2 \pi \left( \frac{r}{\lambda r_*} - \frac{1}{2} \right) \right\}
\ee
and
\be
\nabla_r \gamma^{KS}_{rr} = \frac{2 \pi (t-a) V^2}{r_* \, t} \sin \left\{ 2 \pi \left( \frac{r}{\lambda r_*} - \frac{1}{2} \right) \right\} \, .
\ee
These result in the non-zero components of $\mu_{abcdef}$ being given by
\begin{eqnarray}
\mu^{ST}_{abcdef} &=& 
\frac{2 \pi^2 (t-a)^2 V^2}{r_*^2 \, t^2}
\delta^r_a \delta^r_b\delta^r_c\delta^r_d\delta^r_e\delta^r_f  \qquad {\rm in \; the\; vacuum\; regions} \\
\mu^{KS}_{abcdef} &=& 
\frac{2 \pi^2 (t-a)^2 V^4}{r_*^2 \, t^2}
\delta^r_a \delta^r_b\delta^r_c\delta^r_d\delta^r_e\delta^r_f  \qquad {\rm in \; the\; dust \; regions} \, ,
\end{eqnarray}
which gives the back-reaction tensor in Equation (\ref{GW5}) as
\be
t^{(0)}_{ab} =0
\ee
for all values of $r$. Once more, there appears to be no back-reaction from the vacuum regions on the homogeneous dust-filled background space-time in the Green \& Wald formalism. This is despite the fact that the back-reaction scalar in the Buchert formalism, for the exact same space-time, is always non-zero. This indicates that the measures of back-reaction in the two diffferent formalisms are quantifying quite different phenomena.\\

\noindent {\it Case V: A Schwarzschild T-region background.} We can consider exactly the same set-up, still within the Green \& Wald formalim, but this time choose our ``background'' to be given by the vacuum Schwarzschild T-region. In this case the dust-filled regions are treated as pertubations of the vacuum background. In this case, in the vacuum regions we choose our parameterization of $A(r)$ and $B(r)$ such that
\be
A^{ST} = 1+ \frac{\lambda}{2} \left[ \cos \left\{ 2 \pi \left( \frac{r}{\lambda r_*} - \frac{1}{2} \right) \right\} -1 \right]
\ee
\be
B^{ST}= 0 \, ,
\ee
while in the dust-filled regions we choose
\be
A^{KS}= 1-\lambda
\ee
\be
B^{KS}=\frac{\lambda}{2} \left[ \cos \left\{ 2 \pi \left( \frac{r}{\lambda r_*} - \frac{1}{2} \right) \right\} +1 \right] \, .
\ee
These two functions have the same form as given in Figure \ref{c1c2b}, again with $A \leftrightarrow c_2$ and $B \leftrightarrow c_1$ so that the Schwarzschild T-region is in $0 <r < \lambda r_*$ and the Kantowsi-Sachs region is in $ \lambda r_*< r < 2\lambda r_*$.

In each of the two regions the perturbation can then again be written as in Equation (\ref{KSSTpert}), which gives the only components of $\nabla_a \gamma_{bc}$ that are not proportional to $\lambda$ or $\lambda^2$ as being
\be
\nabla_r \gamma^{ST}_{rr} = \frac{2 \pi (t-a)}{r_* \, t} \sin \left\{ 2 \pi \left( \frac{r}{\lambda r_*} - \frac{1}{2} \right) \right\}
\ee
and
\be
\nabla_r \gamma^{KS}_{rr} = \frac{2 \pi (t-a) V}{r_* \, t} \sin \left\{ 2 \pi \left( \frac{r}{\lambda r_*} - \frac{1}{2} \right) \right\} \, .
\ee
The non-zero components of $\mu_{abcdef}$ can then all be written as
\be
\mu^{ST}_{abcdef} = 
\frac{2 \pi^2 (t-a)^2}{r_*^2 \, t^2}
\delta^r_a \delta^r_b\delta^r_c\delta^r_d\delta^r_e\delta^r_f  \qquad {\rm in \; the\; vacuum\; regions}
\ee
and
\be
\mu^{KS}_{abcdef} = 
\frac{2 \pi^2 (t-a)^2 V^2}{r_*^2 \, t^2}
\delta^r_a \delta^r_b\delta^r_c\delta^r_d\delta^r_e\delta^r_f  \qquad {\rm in \; the\; dust \; regions} \, ,
\ee
and the back-reaction tensor is
\be
t^{(0)}_{ab} =0
\ee
at all values of $r$. Once more, there is no back-reaction with this choice of background, even though the large-scale expansion of the Schwarzschild T-region is quite different to that of the dust-filled Kantowski-Sachs solution.\\

\noindent {\it Case VI: An intermediate background.} In this case we choose the background to have $A=B=1/2$, so that the background is a Kantowski-Sachs solution with a density of matter that is somewhere between the original vacuum and dust-filled regions. In this case, in the vacuum regions we choose
\be
A^{ST} = \frac{1}{2}+ \frac{\lambda}{2} \cos \left\{ 2 \pi \left( \frac{r}{\lambda r_*} - \frac{1}{2} \right) \right\}
\ee
\be
B^{ST}= \frac{1-\lambda}{2} \, ,
\ee
while in the dust-filled regions we choose
\be
A^{KS}= \frac{1-\lambda}{2}
\ee
\be
B^{KS}=\frac{1}{2}+\frac{\lambda}{2} \cos \left\{ 2 \pi \left( \frac{r}{\lambda r_*} - \frac{1}{2} \right) \right\} \, .
\ee
These are the functions displayed in Figure \ref{c1c2c}, again with $A \leftrightarrow c_2$ and $B \leftrightarrow c_1$.

In each of the two regions the perturbation can then again be written as in Equation (\ref{KSSTpert}), which gives the only components of $\nabla_a \gamma_{bc}$ that are not proportional to $\lambda$ or $\lambda^2$ as being
\be
\nabla_r \gamma^{ST}_{rr} = \frac{\pi (t-a) (1+V)}{r_* \, t} \sin \left\{ 2 \pi \left( \frac{r}{\lambda r_*} - \frac{1}{2} \right) \right\}
\ee
and
\be
\nabla_r \gamma^{KS}_{rr} = \frac{\pi (t-a) V (1+V)}{r_* \, t} \sin \left\{ 2 \pi \left( \frac{r}{\lambda r_*} - \frac{1}{2} \right) \right\}
\ee
the non-zero components of $\mu_{abcdef}$ can then all be written as
\begin{eqnarray}
\hspace{-30pt}
\mu^{ST}_{abcdef} &=& 
\frac{\pi^2 (t-a)^2 (1+V)^2}{2 r_*^2 \, t^2}
\delta^r_a \delta^r_b\delta^r_c\delta^r_d\delta^r_e\delta^r_f  \qquad {\rm in \; the\; vacuum\; regions}
\\
\hspace{-30pt}
\mu^{KS}_{abcdef} &=& 
\frac{\pi^2 (t-a)^2 V^2 (1+V)^2}{2 r_*^2 \, t^2}
\delta^r_a \delta^r_b\delta^r_c\delta^r_d\delta^r_e\delta^r_f  \qquad {\rm in \; the\; dust \; regions} \, .
\end{eqnarray}
This gives
\be
t^{(0)}_{ab} =0
\ee
at all values of $r$. As in all previous cases, there is no back-reaction within the Green \& Wald formalism. This result is true for all of the plane and spherically symmetric space-times with $R'=0$ that we have studied in this section.

At this point we note that there do not appear to be any solutions with $R' \neq 0$ that allow the same physical set-up to be considered (i.e. where we have alternating regions of homogeneous dust and vacuum sandwiched between each other in a repeated periodic way forever). This is not to say that homogeneous dust and vacuum regions cannot be matched together when $R'\neq 0$. Indeed it is quite possible to match Friedmann and Schwarzschild R-regions when $k=1$, and to match vacuum Taub and Friedmann solutions when $k=0$. The difficulty is that when one matches multiple successive bands of dust-dominated and vacuum regions together, one finds that the density of the dust in each region cannot be identical in every region unless the foliation is associated with the rest spaces of non-geodesic obbservers. Such a foliation would be most unnatural, in terms of its expansion, as it would mix together contributions from both the expansion of space and the acceleration of the observers. This is not what is usually meant by the term ``expansion'' in cosmology, and so we omit such solutions from this section.

\section{Inhomogeneous dust models}
\label{idm}

\begin{figure}
\begin{center}
\includegraphics[scale=1]{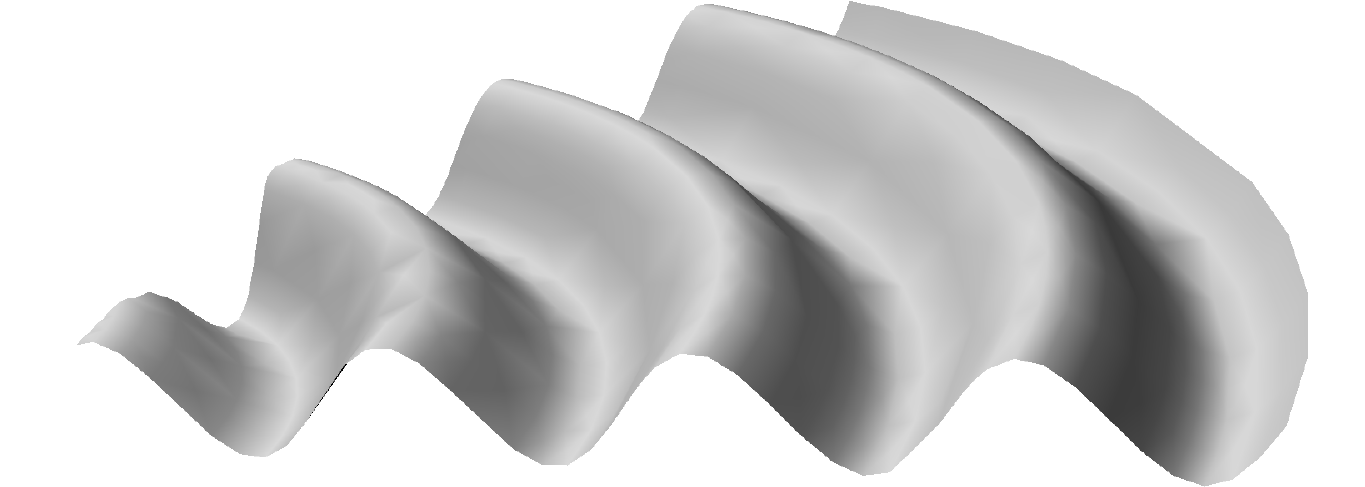}
\caption{An illustration of the situation studied in Section \ref{idm}. The height of the surface is intended to signify the magnitude of the energy density, which is assumed to oscillate around a homogeneous background.}
\label{wave}
\end{center}
\end{figure}

In this section we will examine the possibility of an inhomogeneous energy density that oscillates around a smooth ``background'' value, as illustrated in Figure (\ref{wave}). For this we will use the class of solutions with $R^{\prime} \neq 0$ and $k=1$, which constitute the well-known Lema\^\i tre--Tolman--Bondi (LTB) models. These models exhibit radial dependence in the dust density and other physical and geometric variables, and we will consider how Buchert's and Green \& Wald's averaging formalisms can be applied and compared within them. For a perturbative treatment of inhomogeneous dust sources see Ref. \cite{glod}.

The spherically symmetric LTB models, with $R' \neq 0$ and $k=1$ in Equation (\ref{thing1}), are characterized by the LTB metric:
\begin{equation}
ds^2 = -dt^2+\frac{R'{}^2}{1-K}+R^2\left(d\theta^2+\sin^2\theta\,d\,\phi^2\right),
\label{LTB1b}
\end{equation}
where $R=R(t,r)$ is the ``curvature area distance'' that satisfies the Friedman--like Equation (\ref{Rdiff}), $\rho$ is given by Equation (\ref{thing2}), and $K=K(r)$ determines the spatial curvature. For our current purposes, it turns out to be more convenient to rewrite Equation (\ref{LTB1b}) as the following FLRW--like metric:
\begin{equation} 
ds^2 = -dt^2 + a^2\left[\frac{\Gamma^2\,R'_0{}^2\,dr^2}{1-\bar H_0^2\Omki R_0^2}+R_0^2\left(d\theta^2+\sin^2\theta\,d\,\phi^2\right)\right],
\label{LTB2}
\end{equation}
where
\begin{equation} 
a \equiv \frac{R}{R_0}\qquad {\rm and} \qquad \Gamma \equiv \frac{R'/R}{R'_0/R_0}=1+\frac{a'/a}{R'_0/R_0},
\label{newvars} 
\end{equation}
with Equations (\ref{Rdiff}) and (\ref{thing2}) becoming
\begin{equation} 
\frac{\dot a^2}{\bar H_0^2}=\frac{\Ommi}{a}-\Omki \qquad {\rm and} \qquad \rho = \frac{\rho_0}{a^3\,\Gamma},
\label{thing2b} 
\end{equation}
where the subscript $0$ denotes evaluation at some ``initial'' time $t=t_0$. This means that the constant $\bar H_0$ can be identified with the FLRW Hubble constant, and that the ``quasilocal'' functions $\Ommi(r)$ and $\Omki(r)$ can be given in terms of the initial density $\rho_0(r)$ and the Ricci scalar of the hypersurface $t=t_0$ ({\it i.e.} $\RR_0(r)$) by
\be
\fl \hspace{0.5cm} \Ommi=\frac{2M}{\bar H_0^2 r^3}=\frac{3}{r^3}\int_0^r{\Ommmi\,r^2\,dr} 
\qquad {\rm and} \qquad
\Omki = \frac{K}{\bar H_0^2 r^2} = \frac{3}{r^3}\int_0^r{\Ommki\,r^2\,dr},
\label{initvars1}  
\ee
where we have introduced the following dimensionless forms for the density and spatial curvature scalars: 
\be 
\fl \hspace{0.5cm} \Ommmi\equiv \frac{8\pi\rho_0}{3\bar H_0^2}=\Ommi+\frac{r\,[\Ommi]'}{3},\qquad \Ommki\equiv\frac{\RR_0}{6\bar H_0^2}=\Omki+\frac{r\,[\Omki]'}{3},
\label{initvars2}
\ee
and where we have assumed (as a radial coordinate choice) that $R_0=r$, so that $a_0=\Gamma_0=1$. We note that the functions $\Ommi$ and $\Omki$ can be identified with ``quasi--local'' averages of the initial density and spatial curvature (see References \cite{qaverage1,exactperts,qaverage2,radprofs} for comprehensive discussion on these variables).

The Friedman--like equation (\ref{thing2b}) admits analytic solutions for $a$, which we give in \ref{AppLTB}. The remaining metric function $\Gamma=1+ra'/a$ can be found by implicit differentiation of these solutions, and can be written as
\begin{eqnarray} 
\Gamma = 1+3\,\GG_m\,\ddrho_0-3\,\GG_k\,\ddKK_0,
\label{Gammadef}
\end{eqnarray}
where $\GG_m,\,\GG_k$ are given in \ref{AppLTB}, and where
\ba
\fl \hspace{0.5cm} \ddrho_0 = \frac{r\,[\Ommi]'}{3\Ommi} =\frac{\Omega_0^{(m)}-\Ommi}{\Ommi}
\qquad {\rm and} \qquad
\ddKK_0 = \frac{r\,[\Omki]'}{3\Omki}=\frac{\Omega_0^{(k)}-\Omki}{\Omki},
\label{deltas} 
\ea
with $(\Omega_0^{(m)},\,\Omega_0^{(k)})$ related to $(\rho_0,\,\RR_0 )$ by Equations (\ref{initvars1})--(\ref{initvars2}).  The reader may notice that in this formalism any LTB model becomes completely determined by specifying $\Omega_0^{(m)}$ and $\Omega_0^{(k)}$ as dimensionless initial conditions. The time dependent metric function $a$ then depends on $r$ only through $\Ommi$ and $\Omki$, while the radial dependence of $\Gamma$ involves these initial functions and their gradients, $[\Ommi]'$ and $[\Omki]'$. 

\subsection {Buchert Averages}

Let us first consider the Buchert average of these models. A sufficient condition for a non-zero backreaction term $\QQ_{\DD},$ in arbitrary spherical comoving domains $\DD$ of generic LTB models, is known to be given by \cite{sussBR}
\be 
\QQ_\DD=\langle\hbox{Q}\rangle_\DD\ne 0,
\label{BR4LTB}
\ee
where $\hbox{Q}\equiv \left[\HH-\HH_p\right]^2-\left[\HH-\HH_q\right]^2\ne 0$, and where $\HH\equiv\Theta/3=\dot a/a+\dot\Gamma/(3\Gamma)$ is the Hubble scalar and $\HH_q,\,\HH_p$ are the averaged functions
\be 
\fl \hspace{0.5cm}
\HH_q(t,r) = \frac{\int_\DD{\HH\dd \VV_q}}{\VV_q}, 
\qquad 
\HH_p(t,r) = \frac{\int_\DD{\HH \dd \VV_p}}{\VV_p}
\qquad {\rm and} \qquad
\dd\VV_q=\FF\dd\VV_p,
\label{HHpq}
\ee 
where $\FF=\left[1-K\right]^{1/2}$ and $\dd\VV_p=\sqrt{\hbox{det}(h_{ab})}\dd r\,\dd\theta\,\dd\phi$ is the proper spatial volume element associated with the LTB metric ($h_{ab}=u_au_b+g_{ab}$). It is known that $\hbox{Q}=0\,\,\Rightarrow\,\,\QQ_\DD=0$ for all spatially flat (``parabolic'') LTB models with $\Omki=0$. Hence, we will only consider open (``hyperbolic'') and closed (``elliptic'') models with $\Omki\ne 0$, which are introduced in \ref{appopen}.\\

\noindent 
{\it Open LTB models:}
\\

\noindent
We consider spherical comoving domains $\DD$ bounded by arbitrary fixed $r$.  For the metric in Equation (\ref{LTB2}), with $R_0=r$, the integrals  in Equation (\ref{HHpq}) 
%take the form
%\ba 
%\HH_q(t,r) &=& \frac{\int_\DD{\HH\dd \VV_q}}{\VV_q}
%\qquad {\rm and} \qquad
%\HH_p(t,r)  =\frac{\int_\DD{\HH \dd \VV_p}}{\VV_p} 
%\label{HHpqInt}
%\ea
%and the 
have spatial volumes that can be found to be
\ba
\VV_q &=& \int_\DD{\FF\,\dd\VV_p}=4\pi\int_0^r{a^3\Gamma\dd x}=\frac{4\pi}{3}R^3,\label{VVqInt}\\ 
\VV_p &=& \int_\DD{\dd\VV_p}=4\pi\int_0^r{\frac{a^3\Gamma\dd x}{\FF}}, 
\label{VVpInt}
\ea
where $\dd\VV_p=a^3\Gamma\FF^{-1}\dd r\,\dd\theta\,\dd\phi$ with $\FF=\left[1-\bar H_0^2\Omega_{q0}^k r^2\right]^{1/2}$ and $R=a\,r$.  In general, the back-reaction scalar calculated in such domains will be non-zero, such that $\hbox{Q}(r)\neq 0$ for finite $r$, and will have a magnitude that is a complicated function of both $r$ and the radial profile of the inhomogeneities. 

This complexity is an inevitable consequence of the symmetry and topology of space in such situations, which can no longer be broken up into repeated regions that are identical to each other up to spatial translations. This in turn means that, unlike the locally-homogeneous cases studied in Section \ref{hdm}, there is no simple closed form expression that can be presented for the back-reaction of the global space-time. However, if one were to consider models that converge to FLRW with $\Omega_{q0}^k\ne 0$, then there are simple solutions for back-reaction within averaging domains that encompass the whole time slice (i.e. that have $r\to \infty$). In this case $\HH_q/\HH_p\to 1$ holds as for all open, as shown in Reference \cite{sussBR}. We then have then that backreaction vanishes for such domains in these models, as $\hbox{Q}(r)\to 0$ and thus $\QQ_\DD(r)\to 0$ as $r\to\infty$.\\

\noindent
{\it Closed LTB models:} \\

\noindent
These models are all elliptic, with $\Omki>0$ converging to $\barOmki>0$ (see \ref{appopen} for a discussion of how to avoid thin surface layers in this case). While $\HH_q$ and $\HH_p$ are also defined by Equation (\ref{HHpq}), the individual integrals are qualitatively different. The volume element $\dd\VV_p$ for the metric in this case takes the form
\be
\fl \quad
\dd\VV_p =\frac{a^3\Gamma\,|\cos(\sqrt{k_0}r)|\dd r\,\dd\theta\,\dd\phi}{\FF},
\qquad {\rm where} \qquad
\FF=\left[1-\frac{\Omki}{\barOmki}\sin^2(\sqrt{k_0}r)\right]^{1/2},
\ee
with $k_0$ given by Equation (\ref{R0closed}), and the integrals being evaluated from the first symmetry centre at $r=0$ to the second one at $r=r_c=\pi/\sqrt{k_0}$. The absolute value of the cosine in this volume element is needed to keep $\dd\VV_p\geq 0$. The integrals equivalent to Equations (\ref{VVqInt}) and (\ref{VVpInt}) are then
\be 
\fl \quad \VV_q = \frac{4\pi\,a^3 R_0^3}{3}=\frac{4\pi\,a^3}{3} \frac{\sin^3(\sqrt{k_0} r)}{k_0^{3/2}}
\qquad {\rm and} \qquad
\VV_p=4\pi\int_0^r{\frac{a^3\Gamma |\cos(\sqrt{k_0}r)|\dd r}{\FF}}.
\ee 
Thus, while $\VV_q$ above vanishes at both symmetry centres, $\VV_p$ does not vanish at $r=r_c$. The leading term in series expansions around $r=r_c$ therefore necessarily yields $\VV_q \sim O(r-r_c)^3$, whereas $\VV_p$ has a constant leading term $\VV_p(t,r_c)$. Likewise, the leading-order terms of the integrals in the numerators are different, and thus we have in general $\HH_q\ne \HH_p$ when evaluated at $r=r_c$ for all $t$, which implies  $\hbox{Q}(t,r_c)\ne 0$ so that $\QQ_\DD\ne 0$ for domains $\DD$ encompassing the whole time slice. Therefore, in this case Buchert's formalism yields non-zero backreaction.

\subsection {Green \& Wald Averages}

The Green \& Wald formalism is based on expressing the metric coefficients $g_{ab}$ as a one parameter sequence of functions that comply with standard regularity conditions and that --somehow-- converge to a background metric $g_{ab}^{(0)}$ when the wavelength of the fluctuations becomes vanishingly small. In the case of LTB models, the metric functions are determined from a given choice of initial density and spatial curvature, $\Ommmi$ and $\Ommki$. The short wavelength limit then follows from a one--parameter sequence of LTB models characterized by a sequence of functions
\be 
\Ommmi(\lambda,r) \qquad {\rm and} \qquad \Ommki(\lambda,r), \qquad {\rm for} \qquad \lambda> 0,
\label{GWseq1}
\ee
which yields the sequence of initial value functions $\Ommi(\lambda,r)$ and $\Omki(\lambda,r)$ (and their gradients) as well as $\ddrho_0(\lambda,r)$ and $\ddKK_0(\lambda,r)$) from Equations (\ref{initvars1}) and (\ref{deltas}). The Green \& Wald formalism requires that in the limit $\lambda\to 0$ the metric coefficients associated with the sequence of models should converge to their (yet unspecified) background values. The sequence of functions in Equation (\ref{GWseq1}), and thus the auxiliary functions obtained from Equations (\ref{initvars1}) and (\ref{deltas}), must be constructed in terms of suitable oscillating functions with periods depending on $\lambda$, that keep decreasing with $\lambda$ and reaching small values as it tends to zero. For all $\lambda>0$ the functions in the sequence must be smooth and bounded and must comply with centre and regularity, namely that for all $\lambda$ the gradients of $\Ommmi,\,\Ommki,\,\Ommi,\,\Omki$ vanish at $r=0$ and $\ddrho_0(\lambda,0)=\ddKK_0(\lambda,0)=0$ (see \ref{appreg}).

The physical interpretation of this type of short wavelength limit is that of a continuous one--parameter sequence of LTB models in which the radial profiles of the local density and/or spatial curvature fluctuate in an oscillatory manner around some (yet unspecified) background value that could, in principle, be a unique FLRW dust model characterized by the values $\barOmmi$ and $\barOmki$ (see \ref{appflrw}). The wavelength parameter $\lambda$ provides the scale of these oscillatory fluctuations in terms of the comoving radius, and decreasing values denote decreasing wavelengths and increasing frequency. Demanding that the density and spatial curvature be given by smooth functions that are everywhere bounded, and also oscillatory and periodic, necessarily requires the amplitude of the oscillations to converge to zero as $\lambda\to 0$ at least as fast as the wavelength $\lambda$. In this case the metric of the background model will be smooth in the limit $\lambda\to 0$.  If the amplitude decreases slower than the wavelength, then the LTB metric tends to a distribution in the limit $\lambda\to 0$.

Models with this type of oscillatory energy density have been referred to as ``onion models'' in the literature \cite{onion}, and while they do not describe any real astrophysical structures, they are nonetheless ideal as toy models for investigating the effects of cosmological inhomogeneity in a fully relativistic context. Moreover, it is important to acknowledge that this type of model implies periodic changes of sign of $\ddrho$ and/or $\ddKK$, and thus are bound to develop shell crossings, especially for elliptic models with $\Omki>0$ for which the conditions to avoid these features are very stringent (see \ref{appreg} and Reference \cite{radprofs}).  In what follows we will now discuss the Green \& Wald postulates, as given in Section \ref{secgw}, assuming generic initial value functions for the quantities in Equation (\ref{GWseq1}). We assume only that these functions are bounded, oscillatory and periodic, restricted by smoothness (at least $C^2$) and in fulfillment of conditions for centre regularity and the absence of shell crossings (at least for a finite period of their evolution).\\   

\noindent
{\it Postulate 1:} The fulfilment of this postulate only requires convergence (which we will assume to be uniform) of the sequence of LTB metrics $g_{ab}(\lambda,x^c)$ to a well defined and smooth background metric $g^{(0)}_{ab}$, which needs not be an FLRW metric. There are two relevant metric functions, $a(\lambda,t,r)$ and its radial gradient $\Gamma(\lambda,t,r)$, both of which are obtained from a given choice of $\Ommmi(\lambda,r),\,\Ommki(\lambda,r)$, the auxiliary functions $\Ommi(\lambda,r),\,\Omki(\lambda,r)$ and their gradients $\ddrho_0(\lambda,r),\,\ddKK_0(\lambda,r)$. The two main metric coefficients are (under the choice $R_0=r$),
\be 
\fl \quad g_{rr}(\lambda,t,r) = \frac{a^2\Gamma^2}{1-\bar H_0^2\Omki r^2}=\frac{(a+ra')^2}{1-\bar H_0^2\Omki r^2},
\qquad {\rm and} \qquad
g_{\theta\theta}(\lambda,t,r) = a^2\,r^2.
\ee
Since the metric convergence we are seeking involves the gradients $\ddrho_0,\,\ddKK_0$ of $\Ommi,\,\Omki$, we need to assume that the latter functions (and thus the metric coefficients) are at least $C^1$. The background is then defined generically by the following limits on the sequence of initial value functions:
\ba
\fl \quad  \lim_{\lambda\to 0}\Ommmi(\lambda,r) = \Omega_{b0}^{(m)}\equiv \Ommmi(0,r),\quad 
\lim_{\lambda\to 0}\Ommki(\lambda,r) = \Omega_{b0}^{(k)}\equiv \Ommki(0,r),\label{lim1}\\
\fl \quad \lim_{\lambda\to 0}\Ommi(\lambda,r) = \Omega_{qb0}^{(m)}\equiv \Ommi(0,r),\quad 
\lim_{\lambda\to 0}\Omki(\lambda,r) = \Omega_{qb0}^{(k)}\equiv \Omki(0,r),\label{lim2}\\ 
\fl \quad \lim_{\lambda\to 0}\ddrho_0(\lambda,r) = \ddrho_{b0}\equiv \ddrho_0(0,r),\quad \,\,\,\,
\lim_{\lambda\to 0}\ddKK_0(\lambda,r) = \ddKK_{b0}\equiv \ddKK_0(0,r),\label{lim3}
\ea
which determine the background metric functions $a_b=a(0,t,r),\,\Gamma_b=\Gamma(0,t,r)$. We have then two cases (we provide specific ansatzes further ahead):\\

\noindent
{\it Case I: An FLRW background.} Here we have
\ba 
\fl \quad \Omega_{b0}^{(m)}=\Omega_{qb0}^{(m)} =\barOmmi =\hbox{const.},\quad \Omega_{b0}^{(k)}=\Omega_{qb0}^{(k)} =\barOmki=\hbox{const.},\quad \ddrho_{b0}=\ddKK_{b0}=0,
\label{FLRWb}
\ea
which implies $a_b =\bar a(t)$ and $\Gamma_b = 1$.\\

\noindent
{\it Case II: An LTB background.} Which has
\ba
\fl \quad \Omega_{b0}^{(m)}=\Omega_{b0}^{(m)}(r),\,\,\Omega_{qb0}^{(m)} =\Omega_{qb0}^{(m)}(r),\quad \Omega_{b0}^{(k)}=\Omega_{b0}^{(k)}(r),\,\,\Omega_{qb0}^{(k)} =\Omega_{qb0}^{(k)}(r),\nonumber\\ \fl \quad \ddrho_{b0}=\ddrho_{b0}(r),\,\,\ddKK_{b0}=\ddKK_{b0}(r),
\label{LTBb}
\ea
which implies $a_b =a(0,t,r)$ and $\Gamma_b = \Gamma_b(0,t,r)$.\\

\noindent
As shown by the ans$\ddot{{\rm a}}$tze we provide further ahead, there are many ways to define a sequence of LTB models in the short wavelength regime that exhibit  uniform convergence to these two backgrounds as $\lambda\to 0$.\\  

\noindent
{\it Postulate 2:} If we start by defining the (exact) metric perturbation $\gamma_{ab}=g_{ab}(\lambda)-g^{(0)}_{ab}$, then this postulate requires the existence of a smooth positive function $C_1(x^a)$ such that  
\ba
|\gamma_{\theta\theta}| = |a^2-a_b^2|r^2 <\lambda C_1 
\label{GWLTB2}
\ea
and
\ba
|\gamma_{rr}| = \left|\frac{a^2\Gamma^2}{1-\bar H_0^2\Omki\, r^2}-\frac{a_b^2\Gamma_b^2}{1-\bar H_0^2\bar\Omega_{qb0}^{(k)}\,r^2}\right|<\lambda C_1,
\label{GWLTB2b}
\ea
where $a_b,\,\Gamma_b$ and $\Omega_{qb0}^k$ are the limits of $a(\lambda),\,\Gamma(\lambda),\,\Omki(\lambda)$ as $\lambda\to 0$, given by either Equation (\ref{FLRWb}) or (\ref{LTBb}). The functional form of $a(\lambda)$ follows from the solutions in Equations (\ref{solLTBexp})--(\ref{solLTBcol}), whereas $a_b$ follows from the same solutions but replacing $\Ommi(\lambda),\,\Omki(\lambda)$ with their background values $\Omega_{qb0}^{(m)},\,\Omega_{qb0}^{(k)}$, defined by the limits in Equations (\ref{lim1})--(\ref{lim2}). As we are assuming uniform convergence as $\lambda\to 0$ through these limits,   then $|\Ommi-\Omega_{qb0}^{(m)}|$ and $|\Omki-\bar\Omega_{qb0}^{(k)}|$ are $O(\lambda^2)$ quantities, and thus $|a^2-\bar a^2|$ is at least an $O(\lambda^2)$ quantity, and thus there always exists a function $C_1$ (in fact, a bounded function) that fulfills the inequality in Equation (\ref{GWLTB2}). To prove the inequality in Equation (\ref{GWLTB2b}) we note that $\Gamma$ in Equations (\ref{Gammadef}) and (\ref{parGamma1})--(\ref{parGamma3}) has the form of a linear combination of two terms that are products of a quantity (either $\GG_m$ or $\GG_k$) that depends only on $a$, $\Ommi$ and $\Omki$ (see Equation (\ref{GGdef})). As we are assuming that the limit in Equation (\ref{lim3}) holds, which requires uniform convergence of the gradients $[\Ommi]'$ and $[\Omki]'$, then using the same arguments as in the previous paragraph it is evident that the inequality holds.

Some remarks are necessary for the convergence of the sequence of LTB models to a spatially flat Einstein-de Sitter background. Since for any FLRW background we have $\barOmmi >0$, then it is straightforward to show that $\ddrho_0,\,\ddKK_0\to 0$ implies $\Gamma\to 1$ holds as $\lambda\to 0$, and thus both inequalities in Equations (\ref{GWLTB2})--(\ref{GWLTB2b}) hold for LTB models converging to both open ($\barOmki< 0$) and closed ($\barOmki> 0$) FLRW backgrounds. However, the limit of the fluctuation $\ddKK_0$ is undefined as $\lambda\to 0$ for LTB models that converge to an EdS background. In this case, $\Gamma$ takes the following form for $\lambda\bar H_0\ll 1$:
\be 
\Gamma \approx 1+3\GG_k\,\ddKK_0\approx 1\mp\frac{(\bar a^{5/2}-1)\,r[\Omega_{q0}^k]'}{5\bar a^{3/2}\,\Omega_{q0}^m} \, ,
\label{GammaEdS}
\ee
where $\GG_k$ is given by Equation (\ref{GGdef}), the $\mp$ corresponds to the sign of $-\Omki$, and we have used the fact that $a\to \bar a$ as $\lambda\to 0$ holds because of Equations (\ref{lim1})--(\ref{lim2}). If we consider the form of Equations (\ref{sinusoidal})-(\ref{dels}) in the ansatz that we propose further ahead, we see that in this case we also have $\Gamma\to 1$ as $\lambda\to 0$, and thus that there exists a smooth function $C_1$ such that Equations (\ref{GWLTB2})--(\ref{GWLTB2b}) hold.\\

\noindent
{\it Postulate 3:} To continue with the Green \& Wald formalism we require the components of the tensor ${}^{(0)}\nabla_c\,\gamma_{ab}$, where ${}^{(0)}\nabla_c$ is the covariant derivative defined for the background metric $g_{ab}^{(0)}$. Hence
\ba 
{}^{(0)}\nabla_c\,\gamma_{ab}={}^{(0)}\nabla_c\,g_{ab}(\lambda)=\frac{\partial g_{ab}(\lambda)}{\partial x^c}-{}^{(0)}\Gamma^d_{ac}g_{db}(\lambda)-{}^{(0)}\Gamma^d_{bc}g_{da}(\lambda),
\label{nablagamma}
\ea
where we used the fact that ${}^{(0)}\nabla_c\,g_{ab}^{(0)}=0$ and ${}^{(0)}\Gamma^d_{ac}$ are the Christoffel symbols for the background metric $g_{ab}^{(0)}$. All of these components are given explicitly in \ref{components}. As the derivatives of $a$ are either related to $\Gamma=1+ra'/a$ or can be eliminated in terms of $a$ by means of Equation (\ref{thing2b}), the only non--trivial derivatives contained in $\partial g_{ab}(\lambda)/\partial x^c$ are the derivatives $\Gamma'$ and $\dot\Gamma$ which are second derivatives of $a$:

\ba
 \hspace{-50pt}
\Gamma' &=&\frac{3 a}{r}\left[\frac{\partial \GG_m}{\partial a}\ddrho_0+\frac{\partial \GG_k}{\partial a}\ddKK_0 \right]\,(\Gamma-1)+\frac{9}{r}\left[\Omega_{q0}^m\frac{\partial \GG_m}{\partial \Omega_{q0}^m}\left(\ddrho_0\right)^2
+\Omega_{q0}^k \frac{\partial \GG_k}{\partial \Omega_{q0}^k}\left(\ddKK_0\right)^2\right]\nonumber\\ \hspace{-50pt}
&&+\frac{9}{r}\left[\Omega_{q0}^m \frac{\partial \GG_k}{\partial \Omega_{q0}^m}+\Omega_{q0}^k\frac{\partial \GG_m}{\partial \Omega_{q0}^k}\right]\ddrho_0 \ddKK_0+3\GG_m\left(\ddrho_0\right)'+3\GG_k\left(\ddKK_0\right)',\label{Gammaprime}
\ea

\begin{equation}
\hspace{-50pt}
 \dot\Gamma = 3\left[\frac{\partial \GG_m}{\partial a}\ddrho_0+\frac{\partial \GG_k}{\partial a}\ddKK_0\right]\,\dot a,\label{Gammadot}
\end{equation}
where $\GG_m,\,\GG_k$ are given by Equation (\ref{GGdef}). So far we have assumed in Postulates 1 and 2 that $\Ommi,\,\Omki,\,\ddrho_0,\,\ddKK_0$ are at least $C^1$ functions converging to their background values as $\lambda\to 0$, hence all terms involving $\GG_m,\,\GG_k$ and their derivatives are $C^1$ functions of $\{a,\Omega_{q0}^m,\Omega_{q0}^k\}$ that converge to smooth functions of $\{\bar a,\bar\Omega_0^m,\bar\Omega_0^k\}$ in this limit. However, we have not made any assumption on the gradients $[\ddrho_0]'$ and $[\ddKK_0]'$ that appear in $\Gamma'$ above. As Postulate 3 does not require that the limit of ${}^{(0)}\nabla_c\,\gamma_{ab}$ to be well defined as $\lambda\to 0$ (only that it can be bounded by a positive smooth function $C_2$), we have then the following two possibilities: 

\begin{itemize} 
\item[(i)] {\it The trivial case:} If $[\ddrho_0]'$ and $[\ddKK_0]'$ (which appear in $\Gamma'$ in Equation (\ref{Gammaprime})) converge to their smooth background values as $\lambda\to 0$, then it is straightforward to show that the fulfillment of the previous postulates (uniform convergence of $g_{ab}(\lambda)$ to $g_{ab}^{(0)}$) is a sufficient condition for 
\be 
\lim_{\lambda\to 0} {}^{(0)}\nabla_c\,\gamma_{ab}=0.
\ee 
To prove this result we show that $\partial g_{ab}(\lambda)/\partial x^c$ uniformly converges to $\partial g_{ab}^{(0)}/\partial x^c$: 
\ba 
\fl\frac{\partial g_{ab}^{(0)}}{\partial x^c}&=& \lim_{\lambda\to 0}\frac{\partial g_{ab}(\lambda)}{\partial x^c}=\lim_{\lambda\to 0}\left[\lim_{h\to 0}\frac{g_{ab}(\lambda,x^c+h)-g_{ab}(\lambda,x^c)}{h}\right]\\
\fl&=& \lim_{h\to 0}\left[\lim_{\lambda\to 0}\frac{g_{ab}(\lambda,x^c+h)-g_{ab}(\lambda,x^c)}{h}\right] =\lim_{h\to 0}\left[\frac{g_{ab}(0,x^c+h)-g_{ab}(0,x^c)}{h}\right] \nonumber
\ea
which implies
\ba
\lim_{\lambda\to 0} {}^{(0)}\nabla_c\,\gamma_{ab}=\lim_{\lambda\to 0} {}^{(0)}\nabla_c\,g_{ab}^{(0)}=0. 
\ea 
In order to probe this result with the LTB models as a background, we consider the fact that the procedure is wholly analogous to that of an FLRW background. For the latter background we have $[\ddrho_0]',\,[\ddKK_0]'\to 0$ as $\lambda\to 0$, and thus from the non-zero components of ${}^{(0)}\nabla_c\,\gamma_{ab}$ provided explicitly in Equations (\ref{comp1})-(\ref{comp8}), we obtain (component by component) ${}^{(0)}\nabla_c\,\gamma_{ab}=0$ in the limit $\lambda\to 0$.
\item[(ii)] {\it The non-trivial case:} Convergence of $g_{ab}(\lambda)$ to a smooth background $g_{ab}^{(0)}(0)$ as $\lambda\to 0$ only requires convergence of $\Ommmi,\,\Ommki,\,\Ommi,\,\Omki,\,\ddrho_0,\,\ddKK_0$ in this limit, but not of the radial gradients $[\ddrho_0]'$ and $[\ddKK_0]'$. If the latter do not converge to a smooth background value as $\lambda\to 0$, then at least one component of $\partial g_{ab}(\lambda)/\partial x^c$ must not be trivially zero in this limit. From Equations (\ref{comp1})-(\ref{comp8}) we then have 
\be 
\lim_{\lambda\to 0} {}^{(0)}\nabla_c\,\gamma_{ab} =\KK\,\delta^r_a\delta^r_b\delta^r_c,\quad {\rm where} \quad \KK=\lim_{\lambda\to 0}\left[{}^{(0)}\nabla_c\,\gamma_{ab}\right]_{rrr},
\label{nablagamma3}
\ee
and where the ${rrr}$ component above is bounded but (we assume) is not a smooth function as $\lambda\to 0$. We note that this is not necessarily a crucial impediment for the Green \& Wald formalism, as Postulate 3 only requires boundedness, and the limit $\lambda\to 0$ in Postulate 4 is a weak limit.
\end{itemize}

\noindent
{\it Postulate 4:} The tensor ${}^{(0)}\nabla_a \gamma_{cd} {}^{(0)}\nabla_b \gamma_{ef}$ that is used to construct $\mu_{abcedef}$ in Equation (\ref{GW4}) is (by construction) a purely algebraic extension of the tensor ${}^{(0)}\nabla_c\,\gamma_{ab}(x^d,\lambda)$ examined in Postulate 3. Its components are quadratic combinations of the components of this latter tensor. In case (i), the trivial case discussed above, the function inside the limit in Equation (\ref{GW4}) trivially tends to zero as $\lambda\to 0$, which yields as $\lambda\to 0$ for every smooth tensor field $f^{abcbde}$ a trivially vanishing $\mu_{abcedef}$ (as a strong limit implies a weak limit, though the converse is not true). As a consequence, Postulate 4 holds with $t_{ab}^{(0)}=0$ and $T_{ab}=T^{(0)}_{ab}=\bar \rho u_a u_b$, with $\rho=\bar \rho(t)$ if we assume an FLRW background. This is consistent with the findings in Ref. \cite{controversy1}, who considered the consequences of uniform boundedness of ${}^{(0)}\nabla_a \gamma_{cd} {}^{(0)}\nabla_b \gamma_{ef}$, and concluded that $\mu_{abcedef}$ vanishes in such cases.

For the non--trivial case (ii) above, we can assume for the initial value functions the following generic functional dependence compatible with a short wavelength regime, and only restricted by compatibility with Postulates 1, 2 and 3: 
\be 
\fl\quad \Ommmi = \Omega_{b0}^{(m)}+ \hat\lambda^\alpha C^{(m)}(u),\qquad \Ommki =  \Omega_{b0}^{(k)}+ \hat\lambda^\beta C^{(k)}(u),\qquad u\equiv \frac{r}{\lambda},
\label{sinusoidal}
\ee 
where $C^{(m)},\,C^{(k)}$ are suitable smooth sinusoidal functions that are bounded in the limit $\lambda\to 0$ (to comply with previous postulates), $\hat\lambda=\bar H_0\,\lambda$ and $\alpha,\beta>0$. For an FLRW background so that $\Omega_{b0}^{(m)}=\bar\Ommmi$ and $\Omega_{b0}^{(k)}=\bar\Ommki$ (the case of an LTB background is analogous), we have
\ba \fl \Omega_{q0}^{(A)} = \bar\Omega_0^{(A)}+\frac{3\hat\lambda^n}{r^3}\int{C^{(A)}\,r^2 \dd r}=  \Omega_0^{(A)}-\frac{\bar H_0\hat\lambda^{n-1}}{r^3}I^{(A)},\label{Omms}\\
\fl  \delta_0^{(A)}= \frac{r [\Omega_{q0}^{(A)}]'}{3 \Omega_{q0}^{(A)}}=\frac{\bar H_0\hat\lambda^{n-1} I^{(A)}}{ \Omega_0^{(A)} r^3-\bar H_0\hat\lambda^{n-1} I^{(A)}},\label{dels}\\
\fl \left[\delta_0^{(A)}\right]'= \frac{r^2\bar H_0\hat\lambda^{n-1}\left[(\Omega_0^{(A)} r^3-\bar H_0\hat\lambda^{n-1}I^{(A)})\,rC^{(A)}_{,u}-3\Omega_0^{(A)} I^{(A)}\right]}{\left[\Omega_0^{(A)} r^3-\hat\lambda^{n-1}I^{(A)}\right]^2},\label{ddels}\ea 
where $A$ and $n$ stand generically for $A=m,\,k$ and $n=\alpha,\,\beta$, with $C^{(A)}_{,u}$ and $I^{(A)}$ defined by
\ba I^{(A)}=\int{r^3 C^{(A)}_{,u}\,\dd r}, \qquad C^{(A)}_{,u}=\frac{\dd C^{(A)}}{\dd u} = \frac{1}{r^3}\left[I^{(A)}\right]'.\label{delvars}\ea 
In order to comply with Postulates 1, 2 and 3, the functions $C^{(A)}(u)$,  and the integrals $I^{(A)}(u)$ must be (at least) bounded in the limit $\lambda\to 0$. From Equations (\ref{sinusoidal})--(\ref{delvars}), the trivial case follows if $n>1$, whereas the non--trivial case follows if $n=1$, since $0<n<1$ violates Postulates 1 and 2.  Our non-trivial case extends the findings of Ref. \cite{controversy1} to a situation in which one of the components ${}^{(0)}\nabla_a \gamma_{cd}$ becomes a distribution, in the example space-times we consider.

For the non--trivial case we need to evaluate the following weak limit: 
\ba \mu_{abcedef} =\hbox{w-}\lim_{\lambda\to 0} \left[{}^{(0)}\nabla_a\,\gamma_{cd} {}^{(0)}\nabla_b\,\gamma_{ef}\right],\label{wlim}  \ea
under the assumption that $n=1$ in all the variables in Equations (\ref{sinusoidal})--(\ref{delvars}). However, it is worth looking first at the strong limit of the same quantity, component by component (see \ref{components}). Considering Equation (\ref{nablagamma3}), we have 
\ba 
\fl \quad \lim_{\lambda\to 0} \left[{}^{(0)}\nabla_a\,\gamma_{cd} {}^{(0)}\nabla_b\,\gamma_{ef}\right] = \KK^2\,\delta^r_a\delta^r_b\delta^r_c\delta^r_d\delta^r_e\delta^r_f,
\ea
where
\ba
\fl \quad \KK^2=\lim_{\lambda\to 0}\left[{}^{(0)}\nabla_a\,\gamma_{cd} {}^{(0)}\nabla_b\,\gamma_{ef}\right]_{rrrrrr}=\lim_{\lambda\to 0}\left[{}^{(0)}\nabla_c\,\gamma_{ab}\right]_{rrr}\times \lim_{\lambda\to 0}\left[{}^{(0)}\nabla_b\,\gamma_{ef}\right]_{rrr},\label{sixr}
\ea
as only the $(rrrrrr)$ component contains terms proportional to the gradients $[\delta_0^{(A)}]'$, that we are assuming not converging to a smooth limit as $\lambda\to 0$ (see Equation (\ref{comp1})). Since the strong limit for all components of ${}^{(0)}\nabla_a\,\gamma_{cd} {}^{(0)}\nabla_b\,\gamma_{ef}$ is zero, save for the $(rrrrrr)$ component in Equation (\ref{sixr}), and as the strong limit implies the weak limit, the only non-zero components of the weak limit in Equation (\ref{wlim}) must be the $(rrrrrr)$ component. Therefore, for whatever form $\KK^2$ might take in Equation (\ref{sixr}) (and we are not assuming it to be a smooth function), we can write
\be 
\fl \quad \mu_{abcedef} = \hbox{w-}\KK^2\,\delta^r_a\delta^r_b\delta^r_c\delta^r_d\delta^r_e\delta^r_f, \qquad \hbox{w-}\KK^2=\hbox{w-}\lim_{\lambda\to 0}\left[{}^{(0)}\nabla_a\,\gamma_{cd} {}^{(0)}\nabla_b\,\gamma_{ef}\right]_{rrrrrr}\label{mufinal} 
\ee
Now, regardless of the form that $\hbox{w-}\KK^2$ might take, it is straightforward to show that substitution of Equation (\ref{mufinal}) into the definition of $t_{ab}^{(0)}$ in Equation (\ref{GW7}) yields $t_{ab}^{(0)}=0$. Hence, Postulate 4 also holds in the non--trivial case with zero backreaction, with $T_{cd}=T^{(0)}_{cd}=\rho_b u_c u_d$, and with $\rho_b$ the background density associated with $g_{cd}^{(0)}$.\\ 
        
\noindent
{\it A convenient ansatz:} It is useful to further explore the Green \& Wald formalism by means of a more concrete forms of the generic ansatz given in Equations (\ref{sinusoidal})--(\ref{delvars}) for the initial value functions and fluctuations around an FLRW background (generalization to an LTB background is straightforward). Consider the following specific forms  for the sinusoidal functions $C^{(A)}$ and the integrals $I^{(A)}$ in (\ref{sinusoidal})--(\ref{delvars})
\ba
\fl \quad C^{(m)} =\sin ^2 u,\qquad 
I^{(m)} =\frac{\lambda^4}{4}\left[(3-2u^2)u\cos 2u-\frac32(1-2u^2)\sin 2u\right],\label{CIm}\\
\fl \quad C^{(k)} = \cos^4 u,\qquad I^{(k)} = -\frac{\lambda^4}{64}\left[8u(3-8u^2)\cos^4 u -6(1-8u^2)\cos^3 u\sin u\right.\nonumber\\
\fl \qquad \qquad\qquad\qquad\qquad\qquad\left.+72u\cos^2 u-9(5-8u^2)\cos u\sin u-3u(15-8u^2)\right],\label{CIk}
\ea
with the parameters in Equation (\ref{sinusoidal}) selected as $\alpha=\beta=n$ and an FLRW background given by $\Omega_{b0}^{(m)}=\barOmmi$ and $\Omega_{b0}^{(k)}=\barOmki=\barOmmi-1$. Notice that the integrals $I^{(m)},\,I^{(k)}\to 0$ as $\lambda\to 0$, but $C^{(m)}(u),\,C^{(k)}(u)$ and their derivatives behave in this limit like wildly oscillating sinusoidal function of the form $\sim \sin(r/\lambda)$, and thus must be treated as distributions in this limit. Therefore, to have all the initial value functions in Equations (\ref{sinusoidal})--(\ref{delvars}) converging to smooth functions as $\lambda\to 0$ it is necessary to choose $n>1$, which leads to the trivial case for an FLRW background. For the choice $n=1$, we have the non--trivial case in which $\Omega_0^{(A)},\,\Omega_{q0}^{(A)}\to \bar\Omega^{(A)}$ and $\delta_0^{(A)}\to 0$ as $\lambda\to 0$ (thus complying with Postulates 1, 2 and 3), with $[\delta_0^{(A)}]'$ in Equation (\ref{ddels}) bounded but not converging to smooth functions because of the terms $C^{(m)}_{,u}$ and $C^{(k)}_{,u}$ that tend to distributions that must be evaluated through the weak limit integral in Equation (\ref{mufinal}). However, as we shown before, as the only non-zero component of $\mu_{abcdef}$ is $(rrrrrr)$, we still have zero backreaction in the Green \& Wald formalism regardless of the form of the final distribution in the weak limit integral.   

\begin{figure}[t!]
\begin{center}
\vspace{3cm}
\includegraphics[width=\textwidth]{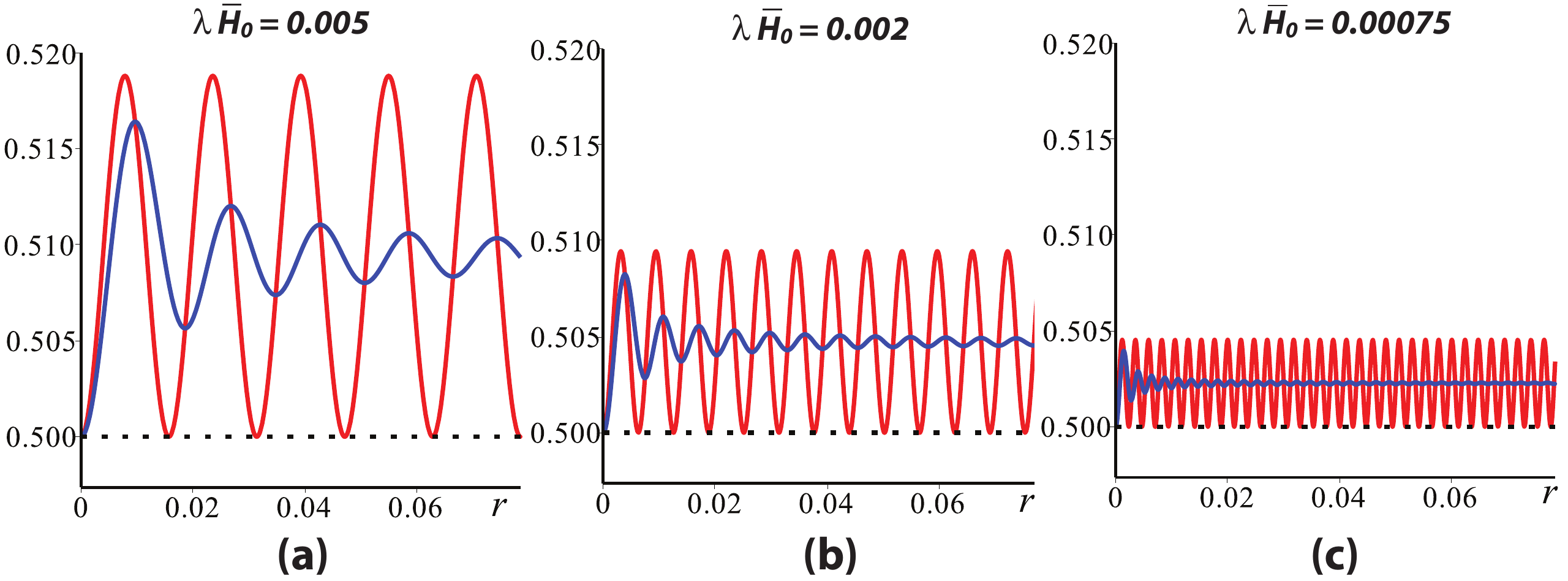}
\caption{The initial value functions $\Ommmi$ (red) and $\Ommi$ (blue)  for various values of $\hat\lambda=\lambda \bar H_0$ for LTB models whose FLRW background is characterised by $\barOmmi=0.5$ and $\barOmmi=\barOmmi-1=-0.5$. These functions are given by Equations (\ref{sinusoidal}) and (\ref{Omms}) for $\alpha=\beta=2$ with $C^{(A)}$ specified by Equations (\ref{CIm})--(\ref{CIk}). Notice how the amplitude decreases to zero as the oscillations converge towards the background value $0.5$}
\label{FigX}
\end{center}
\end{figure}

\begin{figure}[t!]
\begin{center}
\includegraphics[width=\textwidth]{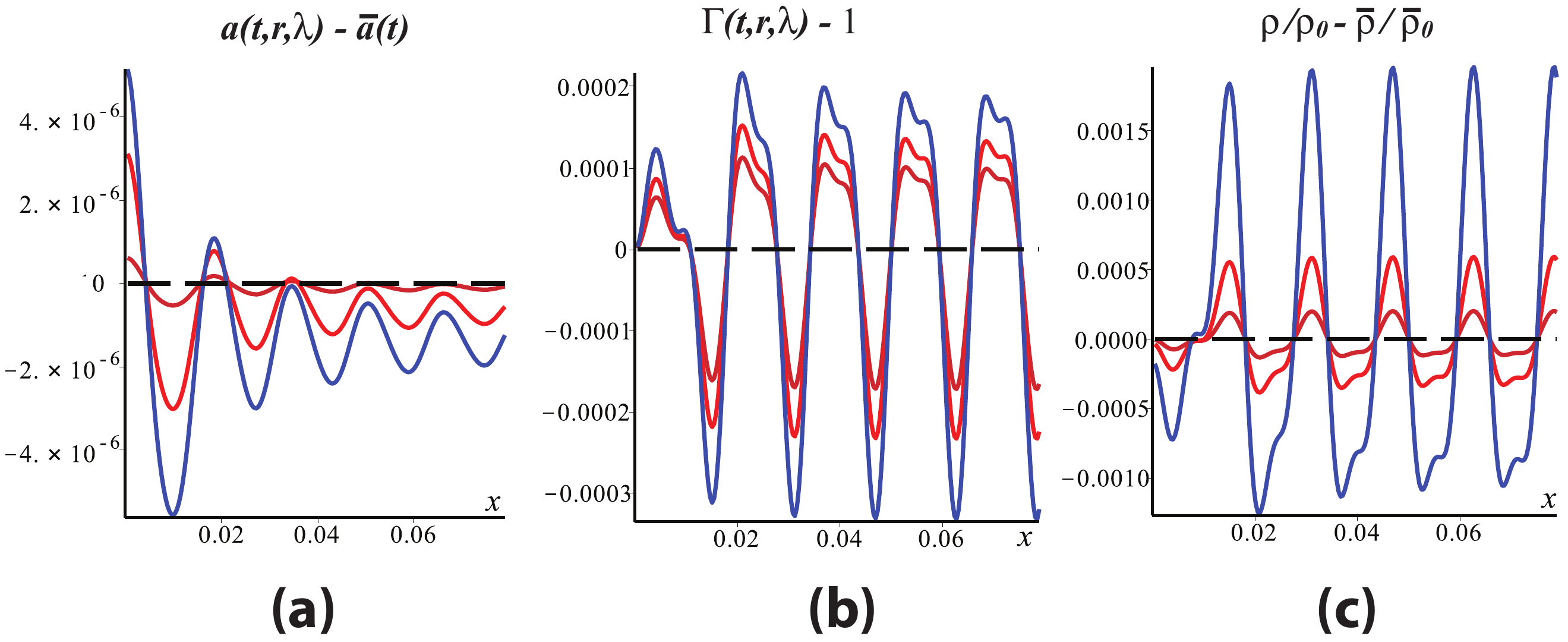}
\caption{The metric functions  $a$ and $\Gamma$ (panels (a) and (b)) and the density ratio $\rho/\rho_0$ (panel (c)), for LTB models whose FLRW background is that of Figure \ref{FigX} with $\lambda \bar H_0=0.005$. The curves correspond to plots as functions of $x=\bar H_0 r$ for fixed times $t=t_0-\delta t/\bar H_0<t_0$ for $\delta t=0.05,\,0.25,\,0.4$ (orange, red, blue).  Notice the oscillatory behavior around the background values (dashed horizontal lines), and how the amplitude of the fluctuations decreases as $t$ grows.}
\label{FigY}
\end{center}
\end{figure}

We depict in Figure \ref{FigX} the functions $\Ommmi$ and $\Ommi$ for various values of $\hat\lambda$ for the selected functional forms in Equations (\ref{CIm})--(\ref{CIk}) of the generic ansatz from Equations (\ref{sinusoidal})--(\ref{delvars}), with $\alpha=\beta=n=2$ and for a negatively curved (hyperbolic) FLRW background $\barOmmi=0.5,\,\barOmki=-0.5$. In Figure \ref{FigY} we display plots of the deviation of the metric functions $a$ and $\Gamma$ from their FLRW values and the density ratio $\rho/\rho_0$, all obtained from the initial value functions plotted in Figure \ref{FigX}, and for various constant values of time. The initial conditions used in the graphic examples of Figures \ref{FigX} and \ref{FigY} depict the oscillations of the initial value functions and metric functions around an FLRW background that characterize the short-wavelength limit used in the Green \& Wald formalism. For these initial conditions it is impossible to avoid shell crossings for the entire time evolution (see Reference \cite{radprofs}), though the conditions to avoid these singularities are much easier to fulfill for hyperbolic models with negative spatial curvature $\Omki<0$, especially with the choice that curvature is more negative where density has a local radial maximum. For elliptic models with positive spatial curvature, $\Omki>0$ (whether open or closed), conditions to avoid shell crossings become too stringent, and regular evolution is only possible for restricted time ranges. Nevertheless, for hyperbolic models these initial conditions are sufficient to get regular oscillatory forms around the FLRW background for the metric coefficients and the density, at least for cosmic times far from the Big Bang. 

\section{Discussion}

We have studied the consequences of applying Buchert's averaging formalism, and Green \& Wald's back-reaction formalism, to a wide array of spherically-symmetric and plane-symmetric dust-filled cosmological models. In Section \ref{hdm} these models were constructed to consist of locally homogeneous dust and vacuum regions sandwiched together, back to back, and repeated over and over again {\it ad infinitum}. In Section \ref{idm} we considered the consequences of applying these two formalisms to generic LTB models that consist of fluctuations in the energy density of dust around a smooth value.

For the locally homogeneous models of Section \ref{hdm}, we found that Buchert's formalism provides a unique and well-defined expression for the kinematical back-reaction of the inhomogeneities on the expansion of finite regions of space. These results display the expected behaviour for backreaction in the limits where the space evolves to towards either homogeneity or strong inhomogeneity, and interpolates smoothly between them. In order to apply Green \& Wald's formalism, on the other hand, required us to identify a ``background''. We first considered the cases in which the dust-filled regions were the ``background'', and the vacuum regions were considered as perturbations. We then considered the situations in which the vacuum regions were the background, and the dust-filled regions were the perturbation. Finally, we choose a background that was somewhere between these two cases, and both the vacuum and dust-filled regions were treated as perturbations. In every case we found that the perturbations had zero effect on the evolution of the background, and hence that the Green \& Wald formalism found no back-reaction.

When studying the generic LTB models with smooth fluctations in the energy density of dust, in Section \ref{idm}, we found similar results. In this case we chose a suitable set of initial value functions to study LTB models that contain sinusoidal waves in the energy density of dust, in the radial direction. We found that global kinematic backreaction vanishes within the Buchert averaging formalism for parabolic (flat) models, but is in general non-zero for hyperbolic (open) models and elliptic (closed) models. We then considered applying the Green \& Wald formalism to similar cases, and carefully assessed the implications of the four postulates of this formalism for the fluctuations. We found, in every case, that the back-reaction term is identically zero in all situations where the postulates are obeyed, such that $t_{ab}^{(0)}=0$. That is, for all LTB models converging to either an FLRW or LTB background in the short-wavelength limit there is zero effect from the short-wavelength fluctuations on the background in the formalism of Green \& Wald.

The fact that the Green \& Wald formalism gives zero back-reaction in all situations studied, including those in which the Buchert kinematic back-reaction is non-zero, arises from the fact that the only non-zero component of the tensor $\mu_{abcdef}$ (used to construct $t_{ab}^{(0)}$) is the $(rrrrrr)$ component, which then yields an identically zero $t_{ab}^{(0)}$ from its definition in Equation (\ref{GW7}). This result shows that the two approaches we have compared must be quantifying different phenomena. We interpret this as follows: The Buchert averaging scheme is designed to pick out the large-scale properties of a space-time, and provides a set of quantities in which this behaviour can be understood. These include quantities like the average of the energy density and the expansion rate of a domain of space, as well as the kinematical back-reaction, which quantifies deviations from the expected Friedmann behaviour. The Green \& Wald formalism, on the other hand, does something quite different. The large-scale properties of the space-time appear to be assumed to be readily identifiable as a ``background''. There is no prescription as to how this background should be identified, which suggests that the authors of this formalism do not consider this to be a question that their approach should be expected to supply. Instead, once a background is given, the Green \& Wald formalism gives the consequences of small scale fluctuations on the average field equations that the background must obey. This is a perfectly valid problem to consider, but is quite different to that which the Buchert formalism appears to be designed to address.

In applying the Green \& Wald formalism to known exact space-times, we found that the lack of a prescription for identifying the ``background'' was a severe limitation on understand how the lack of back-reaction should be understood. Take for example the universe constructed from slabs of Einstein-de Sitter and Kasner geometries. Treating the Kasner regions as perturbations to Einstein-de Sitter is quite different to considering regions of Einstein-de Sitter as perturbations of Kasner. In each case the Green \& Wald formalism tells us there is zero back-reaction from the perturbations, but an Einstein-de Sitter space-time is certainly very different to a Kasner space-time. So, if we have a Universe in which both types of geometry exist and cover approximately equal volumes of space (at some reference time), then how should we describe the large-scale expansion? Likewise, an LTB model can be chosen to have a background that is either an FLRW model or another LTB model. These two background cosmologies may be quite different from each other, yet the back-reaction from the Green \& Wald formalism is zero in both cases.  The Green \& Wald approach does not seem to have an answer to these questions, which are surely of fundamental importance to the question of back-reaction and averaging in cosmology.

\begin{appendix}

\section{LTB models}
\label{AppLTB}

The metric of LTB models contain two metric functions $a$ and $\Gamma=1+ra'/a$, which follow from the solutions of Equation (\ref{thing2b}). The latter admits analytic solutions given by the following implicit relations
\ba 
\fl\quad\hbox{expanding layers}\quad \dot a>0,\,\,\epsilon=\pm 1:\quad \bar H_0(t-t_0) = F-F_0,\label{solLTBexp}\\
\fl\quad\hbox{collapsing layers}\quad \dot a<0,\,\,\epsilon=1:\quad \bar H_0(t-t_0) = 2\pi-(F-F_0),\label{solLTBcol}
\ea
where
\ba
\fl\quad  F=\frac{\epsilon}{\beta_{q0}}\left[A(1-\epsilon\alpha_{q0}a)-\sqrt{\alpha_{q0}a}\sqrt{2-\epsilon\alpha_{q0}a}\right]
\qquad {\rm and} \qquad
F_0=F|_{a=a_0=1},\label{Fdef}
\ea
and where
\ba
\fl \quad \alpha_{q0}=\frac{2|\Omki|}{\Ommi}
\qquad {\rm and} \qquad
\beta_{q0}=\frac{2|\Omki|^{3/2}}{\Ommi},\label{LTBpars1}
\ea
where $A=\arccos$ for $\epsilon=1,\,\,\Omki>0$, and $A=\hbox{arccosh}\,\,\,\hbox{for}\,\,\epsilon-1,\,\,\Omki<0$. The metric function $\Gamma$ is obtained by implicit derivative of (\ref{solLTBexp})--(\ref{solLTBcol}), with the result displayed in Equation (\ref{Gammadef}). The functions $\GG_m$ and $\GG_k$ in this equation are given by
\be 
\GG_m \equiv 1-\frac{\HH_q}{\HH_{q0}}-Y_q,\qquad \GG_k \equiv 1-\frac{\HH_q}{\HH_{q0}}-\frac32 Y_q,
\label{GGdef}
\ee 
with $Y_q\equiv \HH_q(t-t_0)$ given explicitly by
\be
Y_q=\frac{\beta_{q0}\left(2-\epsilon\alpha_{q0}a\right)^{1/2}\,\left[F-F_0\right]}{(\alpha_{q0}\,a)^{3/2}}
\label{parGamma1}
\ee
for expanding layers with $\epsilon=\pm 1,\,\,\HH_q>0$, and
\be
Y_q= -\frac{\beta_{q0}\left(2-\alpha_{q0}a\right)^{1/2}\,\left[2\pi-(F-F_0)\right]}{(\alpha_{q0}\,a)^{3/2}}
\label{parGamma2}
\ee
for collapsing layers with $\epsilon=1,\,\,\HH_q<0$. The expressions for $\HH_q\equiv \dot a/a$ are then 
\be
\fl\quad \HH_q = \frac{\bar H_0\left(\Ommi-\Omki\,a\right)^{1/2}}{a^{3/2}},\qquad \HH_{q0}=\HH_q|_{a=1}=\bar H_0\left(\Ommi-\Omki\right)^{1/2},\label{parGamma3}
\ee
where the second expression is being evaluated at $t=t_0$.

\section{Expanding and collapsing, ``Open'' and ``Closed'', LTB models}
\label{appopen}

The LTB models are usually classified in terms of the sign of $\Omki$, and therefore the existence of zeroes of $\dot a$ through Equation (\ref{thing2b})), into the following kinematic classes: ever expanding ``hyperbolic'' ($\Omki<0$) models, ``parabolic'' ($\Omki=0$) models, and re--collapsing ``elliptic'' models ($\Omki>0$). We can also classify them in terms of the homeomorphic class ({\it i.e.} topology) of the the constant time hypersurfaces. In ``open'' models these hypersurfaces are homeomorphic to $\mathbb{R}^3$, while in ``closed'' they are homeomorphic to 3--spheres $\mathbb{S}^3$ (these models admit other topologies as well, but we will not consider them).   

The coordinate choice $R_0=r$ is appropriate for open models that admit a single symmetry centre (see further below), and in such cases proper length along radial rays diverges as $r\to \infty$. However, ``closed'' models admit two symmetry centres and proper length along radial rays is necessarily bounded. For these models a convenient choice of radial coordinate is 
\be 
R_0=\frac{\sin\,\sqrt{k_0}\,r}{\sqrt{k_0}},\qquad {\rm where} \qquad k_0=\bar H_0^2\bar\Omega_0^k.
\label{R0closed}
\ee
This choice transforms the LTB metric into a form that makes it easier to compare with the metric of a closed FLRW model:
\be 
\dd s^2=-\dd t^2 + a^2\left[\frac{\Gamma^2\,\cos^2(\sqrt{k_0}r)\dd r^2}{1-\frac{\Omega_{q0}^k}{\bar\Omega_0^k} \sin^2(\sqrt{k_0}r)}+\frac{\sin^2(\sqrt{k_0}r)}{k_0}\left(\dd\theta^2+\sin^2\theta \dd\phi^2\right)\right].
\label{LTBclosed}
\ee
Regular closed models must comply with the condition $\Ommi>0$ in order to avoid the existence of a thin shell at the timelike hypersurface $\sqrt{k_0}r=\pi$ (the ``equator'')  where $R'_0=0$ (see comprehensive discussion in \cite{22,23}). Hence, these models are all ``elliptic'', and are characterized by solutions with $\epsilon=1$ in Equations (\ref{solLTBexp}), (\ref{solLTBcol}), (\ref{parGamma1}), (\ref{parGamma2}), and (\ref{Gammadef}).

\section{Regularity conditions}
\label{appreg}

Standard regularity conditions for LTB models are based on the following:\\

\noindent
{\it Centre regularity:} The models admit (up to two) regular centres. ``Open'' models admit a centre, {\it i.e.} a worldline (usually marked by $r=0$) such that $R(t,0)=\dot R(t,0)=0$. ``Closed'' models admit 2 centres, and some models exist without any symmetry centres. \\

\noindent
{\it Absence of shell crossings:} These singularities occur if $R'=0$ holds for $R>0$, or equivalently $\Gamma=0$ for $a>0$. Necessary and sufficient conditions to prevent shell crossings can be derived in terms of initial value functions. These conditions can lead to stringent restrictions on the radial density profiles of the models (especially for elliptic models with $\epsilon=1$ and $\Omki>0$).\\    

\noindent
We will consider only models admitting regular centres, whether one (open models) or two (closed models). Regarding shell crossings, we will assume that the models admit a significant range of time evolution that is free from them.

\section{FLRW background}
\label{appflrw}

Open and closed LTB models may admit a dust FLRW background characterized by the scale factor $\bar a(t)$ and dust density $\bar \rho(t)$ (we shall denote all FLRW quantities by an overbar). For open models these are given by
\ba 
ds^2 = -dt^2 +\bar a^2\left[\frac{dr^2}{1-\bar H_0^2\bar\Omega_0^{(k)}r^2}+r^2\left(d\theta^2+\sin^2\theta\,d\phi^2\right)\right],
\label{FLRWmetric}
\ea
where
\ba
\frac{\dot{\bar a}^2}{\bar H_0^2}=\frac{\barOmmi}{\bar a}-\barOmki,\qquad {\rm and} \qquad \bar \rho=\frac{\bar\rho_0}{\bar a^3}.
\label{FLRW2}
\ea
The constants $\bar H_0$, and $\barOmmi=8\pi\bar\rho_0/(3\bar H_0^2)$, and $\barOmki=\barOmmi-1=\bRR_0/(6\bar H_0^2)$ are the present day Hubble and density factors. The necessary and sufficient conditions for this background to be realised follow by demanding that initial functions be only time dependent, such that  
\ba
\fl \quad [\Omega_0^{(m)}]'=[\Ommi]'=0\quad\Rightarrow\quad \Omega_0^{(m)}=\Ommi=\barOmmi\quad\Rightarrow\quad \ddrho_0=0,\label{FLRWlim1}\\
\fl \quad [\Omega_0^{(k)}]'=[\Omki]'=0\quad \,\,\, \Rightarrow\quad \Omega_0^{(k)}=\Omki=\barOmki\quad \,\,\,\, \Rightarrow\quad \ddKK_0=0.
\label{FLRWlim2}
\ea
These conditions mean that Equations (\ref{thing2b}) and (\ref{Gammadef}) yield $a=\bar a(t)$ and $\Gamma=1$, and thus Equation (\ref{LTB2}) tends to (\ref{FLRWmetric}), and Equation (\ref{thing2b}) tends to (\ref{FLRW2}). For closed models, it is straightforward to show that Equations (\ref{FLRWlim1})--(\ref{FLRWlim2}) applied to Equation (\ref{LTBclosed}) yield the metric of a closed dust-filled FLRW model.

\section{Components of tensors in the Green \& Wald formalism}
\label{components}

The non-zero components of the tensor $\bar\nabla_c\,\gamma_{ab}$ for the LTB metric in Equation (\ref{LTB2}) are 
\ba   
\fl  \left[\bar\nabla_c\,\gamma_{ab}\right]_{rrr} &=& g_{rr}'-\frac{2\bar H_0^2\bar\Ommki r}{1-\bar H_0^2\bar\Ommki r^2}\,g_{rr}=
 \frac{2}{r}\left[ \frac{r\Gamma'}{\Gamma}\,\frac{1-\bar H_0^2\Omki r^2}{1-\bar H_0^2\bar\Ommki r^2}+(1-\bar H_0^2\bar\Ommki r^2)\Gamma\right.\nonumber\\
 \fl &{}&\left.+\bar H_0^2\Omki r^2\left(1+\frac32\ddKK_0\right)-1\right]\,g_{rr}, \label{comp1}
\\
 \fl \left[\bar\nabla_c\,\gamma_{ab}\right]_{trr}&=&\left[\bar\nabla_c\,\gamma_{ab}\right]_{rtr}=-\gamma_{rr}\frac{\dot{\bar a}}{\bar a} = \left[\frac{a^2\Gamma}{1-\bar H_0^2\Omki r^2}-\frac{\bar a^2}{1-\bar H_0^2\barOmki r^2}\right]\frac{\dot{\bar a}}{\bar a},\label{comp2}\\
 \fl \left[\bar\nabla_c\,\gamma_{ab}\right]_{\theta\theta r}&=& 2ra^2(\Gamma-1),\qquad \left[\bar\nabla_c\,\gamma_{ab}\right]_{\phi\phi r}=\sin^2\theta\,\left[\bar\nabla_c\,\gamma_{ab}\right]_{\theta\theta r},\label{comp3}\\
\fl  \left[\bar\nabla_c\,\gamma_{ab}\right]_{\theta r\theta}&=&\left[\bar\nabla_c\,g_{ab}\right]_{r\theta\theta}= ra^2\left[1-\frac{(1-\bar H_0^2\barOmki x^2)\,\Gamma}{1-\bar H_0^2\Omki r^2}\right],\label{comp4}\\
 \fl \left[\bar\nabla_c\,\gamma_{ab}\right]_{r\phi\phi}&=&\sin^2\theta\,\left[\bar\nabla_c\,\gamma_{ab}\right]_{r\theta\theta},\label{comp5}\\
 \fl \left[\bar\nabla_c\,\gamma_{ab}\right]_{t\phi\phi}&=& \left[\bar\nabla_c\,\gamma_{ab}\right]_{\phi t\phi}=-\gamma_{\theta\theta}\sin^2\theta\,\frac{\dot{\bar a}}{\bar a}= \left(\bar a^2-a^2\right)r^2\sin^2\theta\,\frac{\dot{\bar a}}{\bar a},\label{comp6}\\
 \fl \left[\bar\nabla_c\,\gamma_{ab}\right]_{rrt}&=& 2g_{rr}\left[\frac{\dot{\bar a}}{\bar a}-\frac{\dot a}{a}-\frac{\dot \Gamma}{\Gamma}\right],\label{comp7}\\
 \fl \left[\bar\nabla_c\,\gamma_{ab}\right]_{\theta\theta t}&=& \frac{2r^2 a}{\bar a}\left(\dot a\,\bar a-a\dot{\bar a}\right),\qquad \left[\bar\nabla_c\,\gamma_{ab}\right]_{\phi\phi t}=\sin^2\theta\,\left[\bar\nabla_c\,\gamma_{ab}\right]_{\theta\theta t},\label{comp8}
\ea
\\

\end{appendix}

\vspace{-0.5cm}
\noindent {\bf Acknowledgements:}  TC is supported by the STFC grants ST/M001202/1 and ST/P000592/1. RAS was supported by the University of London Perren Fund while on sabbatical at Queen Mary University of London, during which time the majority of the work reported here was undertaken. We are grateful to Alan Coley and Jan Ostrowski for helpful discussions.\\

\end{document}